%% file: mbr.tex
\documentclass[iop,apj]{emulateapj}
\usepackage{apjfonts}

\newcommand{\HI}{H\textsc{i}}

\begin{document}

\title{OGLE-ing the Magellanic System: Stellar populations in the Magellanic Bridge}

\author{D.~M.~Skowron\altaffilmark{1}}
\email{dszczyg@astrouw.edu.pl}
\author{A.~M.~Jacyszyn\altaffilmark{1}}
\author{A.~Udalski\altaffilmark{1}}
\author{M.~K.~Szyma\'{n}ski\altaffilmark{1}}
\author{J.~Skowron\altaffilmark{1}}
\author{R.~Poleski\altaffilmark{1,2}}
\author{S.~Koz{\l}owski\altaffilmark{1}}
\author{M.~Kubiak\altaffilmark{1}}
\author{G.~Pietrzy\'{n}ski\altaffilmark{1,3}}
\author{I.~Soszy\'{n}ski\altaffilmark{1}}
\author{P.~Mr\'{o}z\altaffilmark{1}}
\author{P.~Pietrukowicz\altaffilmark{1}}
\author{K.~Ulaczyk\altaffilmark{1}}
\author{\L.~Wyrzykowski\altaffilmark{1,4}}
\altaffiltext{1}{Warsaw University Astronomical Observatory, Aleje Ujazdowskie 4, \mbox{00-478} Warszawa, Poland}
\altaffiltext{2}{Department of Astronomy, The Ohio State University, 140 W. 18th Ave., Columbus, OH 43210, USA}
\altaffiltext{3}{Universidad de Concepci\'on, Departamento de Astronomia, Casilla \mbox{160-C}, Concepci\'on, Chile}
\altaffiltext{4}{Institute of Astronomy, University of Cambridge, Madingley Road, \mbox{CB3 0HA} Cambridge, UK}

\begin{abstract}

\noindent
We report the discovery of a young stellar bridge, that forms a continuous
connection between the Magellanic Clouds. This finding is based on number
density maps for stellar populations found in data gathered by OGLE-IV, that
fully cover over 270 deg$^2$ of the sky in the Magellanic Bridge area. This is
the most extensive optical survey of this region up to date.
We find that the young population is present mainly in the western half of the
MBR, which, together with the newly discovered young population in the eastern
Bridge, form a continuous stream of stars connecting both galaxies along
$\delta\sim-73.5$~deg. The young population distribution is clumped, with one
of the major densities close to the SMC, and the other, fairly isolated and
located approximately mid-way between the Clouds, which we call the OGLE island.
These overdensities are well matched by HI surface density contours, although
the newly found young population in the eastern Bridge is offset by $\sim2$~deg
north from the highest HI density contour.
We observe a continuity of red clump stars between the Magellanic Clouds, which
represent an intermediate-age population. Red clump stars are present mainly in
the southern and central parts of the Magellanic Bridge, below its gaseous part,
and their presence is reflected by a strong deviation from the radial density
profiles of the two galaxies. This may indicate either a tidal stream of stars,
or that the stellar halos of the two galaxies overlap.
On the other hand, we do not observe such an overlap within an intermediate-age
population represented by the top of the red giant branch and the asymptotic
giant branch stars. We also see only minor mixing of the old populations of the
Clouds in the southern part of the Bridge, represented by the lowest part of
the red giant branch.

\end{abstract}


\section{Introduction}
\label{sec:introduction}

\vspace{0.1cm}
\begin{figure*}[ht]
\centerline{\includegraphics[width=17.9cm]{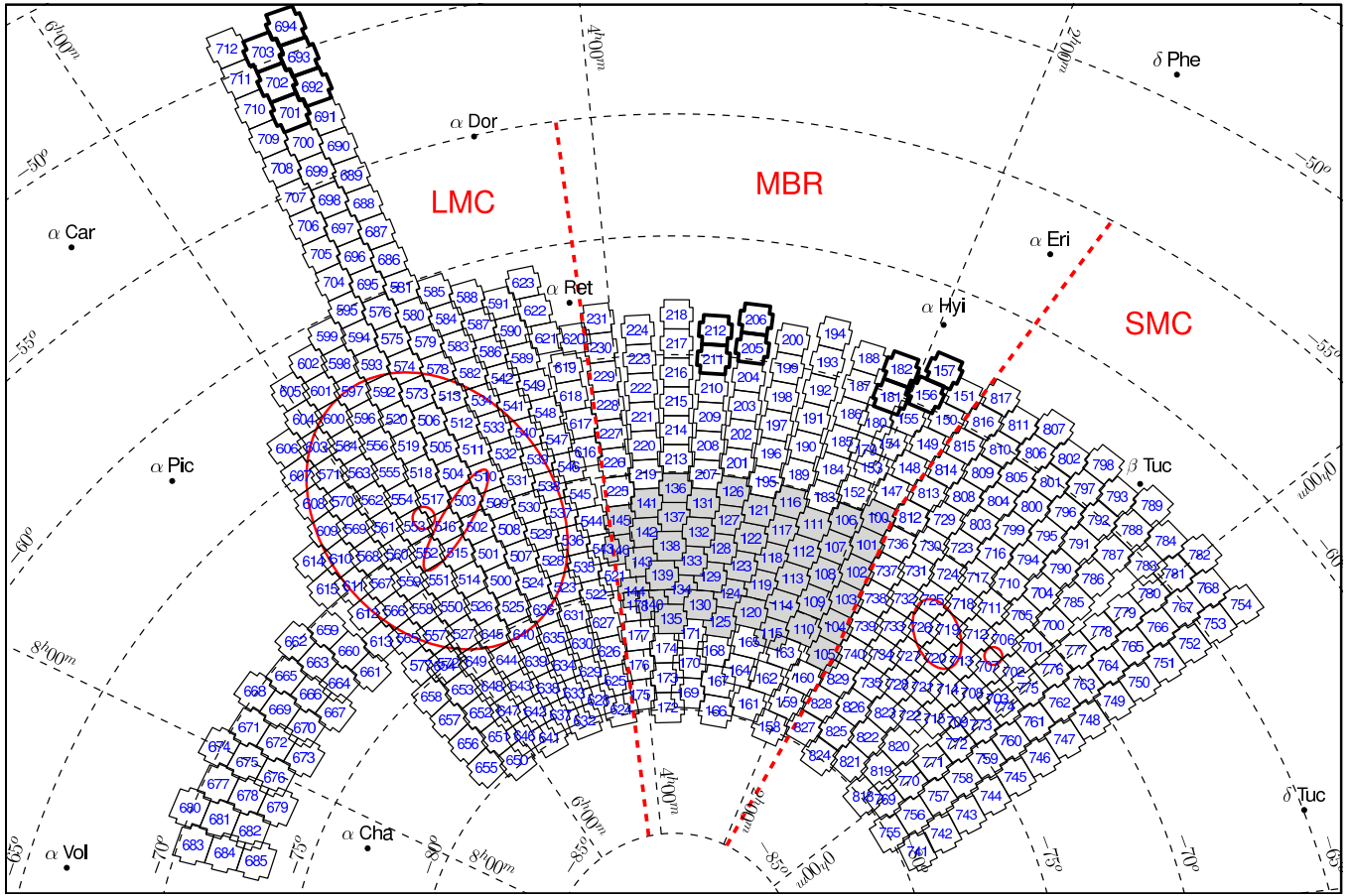}}
\caption{OGLE-IV fields in the Magellanic Clouds region (approximately
600~deg$^2$ of the sky, 1.4~deg$^2$ field area). The Magellanic Bridge region
($\sim$270~deg$^2$) is encompassed by red dashed lines and 47 fields of the
main part of the Magellanic Bridge ($\sim$65~deg$^2$) are shaded with gray.
Fourteen Galactic foreground fields are marked with thick black lines.
Large red ellipses mark the location of the LMC (left) and the SMC (right).
For a more detailed view of the sky coverage see
http://ogle.astrouw.edu.pl/sky/ogle4-fields.html
}
\vspace{0.3cm}
\label{fig:fields}
\end{figure*}
\notetoeditor{(fig:fields) black and white for print, in color only in
the electronic version}

\tabletypesize{\tiny}
\input{table1e.tex}

\tabletypesize{\footnotesize}

The Magellanic Clouds (MCs) comprise of two galaxies: the Large and
the Small Magellanic Cloud (LMC and SMC, respectively), and are the closest to
the Milky Way (MW) pair of interacting galaxies. The Clouds have always been of
special interest to astronomers and they continue to play a significant role
in our understanding of the Universe.

There exists irrefutable evidence that the MCs interact with each other
and with our Galaxy (e.g. \citealt{Besla2012}, \citealt{Diaz2012}):
{\em the Magellanic Bridge} -- a stream of gas and stars between the MCs;
{\em the Magellanic Stream} -- $160$~deg long stream of gas trailing the MCs
on their orbit around the MW;
{\em the Leading Arm} -- a stream of gas leading the MCs on their orbit.
The Magellanic Clouds and all the structures described above are collectively
called the Magellanic System.

The Magellanic Bridge (hereafter MBR) has long been known to contain 
neutral and ionized gas (\citealt{Mathewson1984}, \citealt{Marcelin1985}, 
\citealt{Putman2000}, \citealt{Muller2003}, \citealt{Barger2013}) connecting
the MCs.
It is widely believed that gas present in the MBR has been drawn out of the SMC
through tidal forces, during the most recent encounter of the two galaxies,
that took place 200 Myr ago (\citealt{Mathewson1985}, \citealt{Muller2004}).
Observations of young stars on the SMC side of the Bridge, whose age estimates
are consistent with 200 Myr, support this hypothesis \citep{Irwin1985}.

Early observations of the MBR also revealed a young stellar counterpart
in the  Shapley Wing of the SMC (\citealt{Shapley1940}, \citealt{Meaburn1986},
\citealt{Courtes1995}) and in a number of locations between the Clouds
(\citealt{Irwin1990}, \citealt{Grondin1992}, \citealt{Demers1998}).

\cite{Harris2007} searched for older stellar populations that should have been
drawn out of the SMC by those same tidal forces that drew out gas, in a dozen
fields uniformly sampling the Bridge, most following the ridge-line of neutral
hydrogen ($\delta \sim -74$~deg) and two slightly off. He did not find
any signs of older populations, suggesting, that maybe all stars present in the
MBR formed there, and for some reason tidal forces stripped just pure gas from
the SMC.

Recently, \cite{Bagheri2013} used near infrared public data from 2MASS (Two
Micron All Sky Survey) and WISE (Wide-Field Infrared Survey Explorer) for the
MBR region, spanning 8~deg in declination. They analyzed color-magnitude
and color-color diagrams and found traces of an older population of stars,
estimated to be somewhat between $\sim 400$~Myr and 5~Gyr old. However, the
exact age and the MBR membership of these stars need to be confirmed by further
observations.

Another study of the Bridge region was carried out by \cite{Noel2013}, as
a part of the MAGellanic Inter-Cloud program (MAGIC). They used observations
from two fields between the Clouds, of total area of $1.12$~deg$^2$.
With a synthetic color-magnitude diagram
fitting technique they showed that 28\% of stars in observed regions are
intermediate age and that there are also hints of an older population, that
might have been tidally drawn out of the SMC. However, spectroscopic
observations are needed to confirm or rule out their SMC origin.

Most recently, \cite{Nidever2013} investigated the spatial distribution of
red clump stars in the vicinity of the SMC. They found, that the eastern side
of the SMC has a large line-of-sight depth, which shows a distance bimodality,
with one component being closer to us than the ``systemic'' SMC distance.
The authors argue that this population is a stellar counterpart of the gaseous
MBR, that was stripped from the SMC $\sim$200 Myr ago. However, their data were
not well reproduced by the MCs simulations, which suggests, that the
simulations need to be revised to take into account the larger extent of
stellar components in the MBR.

There is also an on-going survey of the SMC and the MBR performed with the
ESO VLT Survey Telescope, called STEP (the SMC in Time: Evolution of
a Prototype interacting late-type dwarf galaxy). The survey will cover
74~deg$^2$ of the sky with multiple filters to magnitudes fainter than the Main
Sequence turn-off \citep{Ripepi2014}. 

Regarding the Magellanic System simulations, we have to mention here, that
even though there is an agreement, that gaseous features on the Magellanic
System are an effect of some form of interactions between the SMC, LMC and the
MW, there is still an ongoing debate on the nature of these interactions. One
popular scenario is that the gaseous features of the Magellanic System were
created during multiple pericentric passages of the MCs on their orbit around
the MW, either via tidal effects or ram pressure stripping (see the summary
by \citealt{Ruzicka2009}). Another possibility is that the LMC and SMC have
become an interacting pair only recently, with a first close encounter
$\sim$2~Gyr ago and a second $\sim$250~Myr ago.
In the light of recent proper motion measurements based on {\em HST} data
\citep{Kallivayalil2013}, it is highly probable that the MCs are either on
their first infall into our Galaxy or on an eccentric, long period orbit
around the Galaxy \citep{Besla2007}, and have been a bound pair for past
couple of Gyr. Regarding the MBR, \cite{Besla2012} argue, that the internal
kinematics and structure of the Clouds suggest that there was a recent direct
collision ($\sim200$~Myr ago) of the MCs, that produced the Bridge and
triggered star formation within it. This is consistent with simulations of
\cite{Diaz2012}, which show that the MBR was formed due to a strong tidal
interaction $\sim$250~Myr ago.

The Optical Gravitational Lensing
Experiment\footnote{http://ogle.astrouw.edu.pl/} (OGLE) is a long-term
large scale sky survey focused on variability studies of dense stellar
regions. The OGLE project started regular observations in 1992 as one of
the first generation microlensing projects dedicated to detecting and
characterizing microlensing events \citep{Udalski1992}. During its over
22 year history OGLE gradually evolved and conducted numerous projects
that contributed to many fields of modern astrophysics. The current, fourth
phase of the OGLE survey (OGLE-IV) started in March 2010 with the
commissioning of a large new generation 256 Megapixel 32-chip mosaic
camera. The Galactic bulge and disk, the Magellanic Clouds, and the
Magellanic Bridge, including vast areas around them, are the primary
observing targets for the OGLE-IV survey.

In this paper we present density maps of stellar populations in the entire
Magellanic Bridge region, thanks to the unprecedented OGLE-IV coverage. The
maps show, for the first time, the detailed extent of these populations, which
should provide valuable input information for models of past MW and MCs
interactions.

\section{Observations and Data Preparation}

\subsection{OGLE-IV Observations}

Regular monitoring of the selected sky regions by the OGLE survey is
carried out with the 1.3~m Warsaw telescope located at the Las Campanas
Observatory in Chile (operated by the Carnegie Institution for Science)
equipped with the 256 Megapixel 32-chip mosaic camera. The field
of view of the camera is $1.4$ square degrees and a pixel size is
$\sim0\farcs26$. The magnitude range of the standard OGLE survey is
approximately $12-21$~mag in the $I$ band and $12.5-21.5$ mag in the $V$
band.

Figure~\ref{fig:fields} shows OGLE-IV coverage of the Magellanic System.
Shaded MBR fields ($1^{h}40^{m} \lesssim \alpha \lesssim 4^{h}15^{m},
-77^{\circ} \lesssim \delta \lesssim -70^{\circ}$)
have been observed since 2010, while uncolored MBR fields (additional
$\sim 3$~deg south and $\sim 7$~deg north in $\delta$) since 2012.
Number of epochs is on average $280$ in $I$ and $20-36$ in $V$ for the shaded
region and $85$ in $I$ and $3-13$ in $V$ for the rest of the MBR.
In this paper we analyze the entire MBR region, as well as two-field wide
stripes on both the SMC and the LMC side. We also use additional 6 fields
north of the LMC which, together with 8 northern MBR fields, will be used as
a Galactic foreground representation (marked with thick black lines in
Figure~\ref{fig:fields}).
The list of field center coordinates is given in Table~\ref{tab:fields}.

\subsection{Data Reduction}

OGLE observations are reduced on site at the telescope and the photometry is
done in real time.
After standard bias subtraction and flat-fielding (sky flats), images are
processed with the OGLE photometric data pipeline \citep{Udalski2003} that is
based on image subtraction using the ``difference image analysis'' technique
software (DIA, \citealt{Wozniak2000}) adapted to OGLE data \citep{Udalski2008}.
In the first step a reference image for each field is constructed from 3--6
good quality frames. Then each frame is aligned with the reference image for
that field and the reference image is scaled to match the PSF (point spread
function) and background of this frame. In the last step the reference image is
subtracted from the frame, leaving a difference image with flux only from those
objects that either brightened or dimmed.

The database of all sources is created from the reference images with DoPHOT
\citep{Schechter1993} that identifies sources and measures their mean
magnitudes. The light curves are made by adding (or subtracting) the flux from
a difference image to the flux from the reference image. Finally, data are
calibrated to the standard Johnson-Cousins photometric system and the positions
of stars are transformed to equatorial coordinates \citep{Szymanski2011}.

The final photometric database contains 12 million objects (in $I$) in the
entire analyzed Bridge region (both the MBR and LMC/SMC strips), ranging from
25,000 objects per field in sparse areas and 500,000 objects per field in dense
areas close to the LMC.

\subsection{Data cleaning}
\label{sec:datacleaning}

After basic reductions described in the previous subsection, we make several
cuts to the data to obtain a clean, homogeneous sample of the MBR stellar
population. The main source of contamination are detections of spurious sources
located within spikes and halos from saturated stars, ghost reflections from
nearby bright objects, as well as multiple detections on extended objects, etc.
We identify those in two ways: by their light curve scatter and proximity
of similarly looking objects. 

We start by assuming that all sources are spurious detections. For each source
the algorithm counts the number of neighbors within 10, 15, and 20 pixel radii
and remembers their magnitudes. Then a series of conditions are tested which,
if true, change the status of a source to a real detection.
{\it First:} an object has at least two neighbors within $r=20$px and all of
them are fainter than the object, and the sum of the magnitude differences
between this object and each of the neighbors is greater than some
factor $C$ times the number of the neighbors, i.e.
$$ \sum_{i=1}^{n} \left( m_{obj} - m_i  \right) > C \times n$$
In our case $C$ was empirically chosen to be equal to 2/3.
This condition identifies large concentrations of similar-brightness
detections, such as found around bright stars or extended objects, leaving
the brightest of them (e.g. the actual bright star center).
{\it Second:} an object does not have any neighbors within $r=15$px or has
one neighbor within $r=15$px and none within $r=10$px. This condition
ensures that isolated stars are not rejected.
{\it Third:} an object has no more than two close neighbors (within $r=20$px).
If none of the above conditions are true, and object stays marked as a spurious
detection candidate. If at least one condition is true, an object is marked as
real. These criteria are a result of extensive analysis and visual
investigation of many images. The described method works well in uncrowded
fields such as those in the MBR and the false positives (negatives) fraction
is less than 1\%.

In the second step, for each object, we compare measurement errors of
individual observations in the object's light curve as well as the scatter of
observations around the mean magnitude value for this star, with the same
two parameters for a typical star of this magnitude.
This allows for identifying ghost reflections, spikes from bright stars and
other spurious objects, because they typically vary a lot from epoch to epoch,
due to differences in telescope pointing, as well as seeing and background
levels.
Having access to light curves of all objects in our database, we can calculate
a typical rms scatter for a light curve at a given magnitude,
${\rm rms}_{\rm m}$, in the entire magnitude range. Having done that, each
star is assigned a $\Delta_{\rm scatter}$ value which tells how different
its light curve scatter is from ${\rm rms}_{\rm m}$, in units of Gaussian
standard deviations. The same procedure is done for photometric uncertainties
of individual measurements reported by the pipeline, so that real variable
stars are not rejected from the database -- since spurious detection often
do not have regular PSFs, the photometric pipeline tends to assign measurement
errors higher than typical for a star of a given magnitude. Each star is
assigned a $\Delta_{\rm error}$ value which tells us how atypically high
mean uncertainties in its light curve are. When 
$\Delta_{\rm scatter} \times \Delta_{\rm error} > 64$, an object is marked
as spurious. If $\Delta_{\rm scatter} \times \Delta_{\rm error} > 9$, we check
if there are any other objects suspected spurious within 20~px radius,
and if this is the case, we mark the object as spurious.

Lastly, if an object is marked as a spurious detection candidate by both
algorithms, we reject it from the final database. 

\begin{figure}[bt]
\centerline{\includegraphics[width=8.6cm]{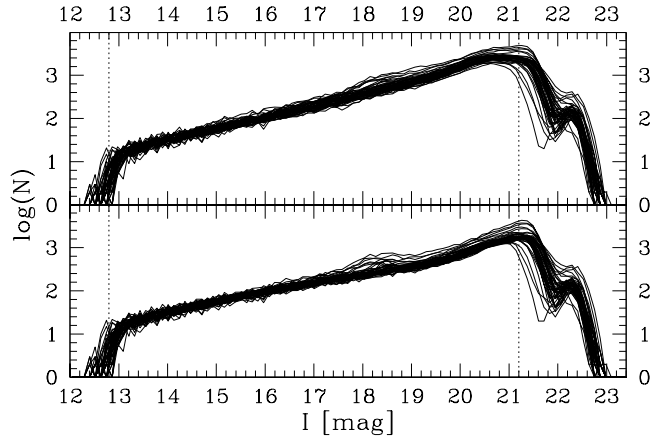}}
\caption{Luminosity functions for 47 fields that cover the main MBR.
The top panel shows data before removing spurious detections and
the bottom panel after the cleaning (for details refer to
Section~\ref{sec:datacleaning}). Dotted lines mark the bright and
faint completeness limits.}
\label{fig:lumfunct}
\end{figure}

\setlength{\tabcolsep}{3pt}
\begin{figure*}[t]
\centering
\begin{tabular}{cccc}
\includegraphics[height=4.9cm]{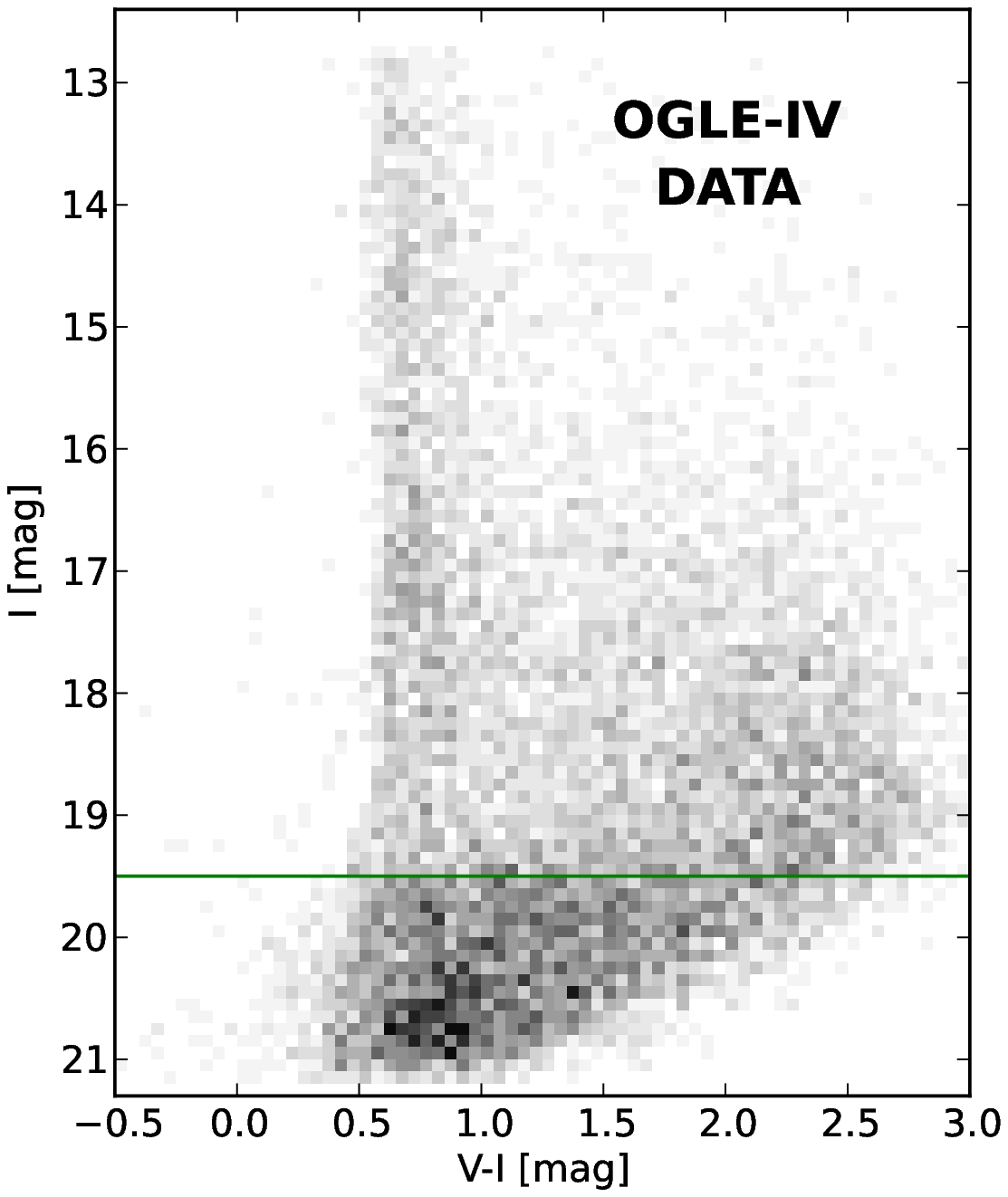} &
\includegraphics[height=4.9cm]{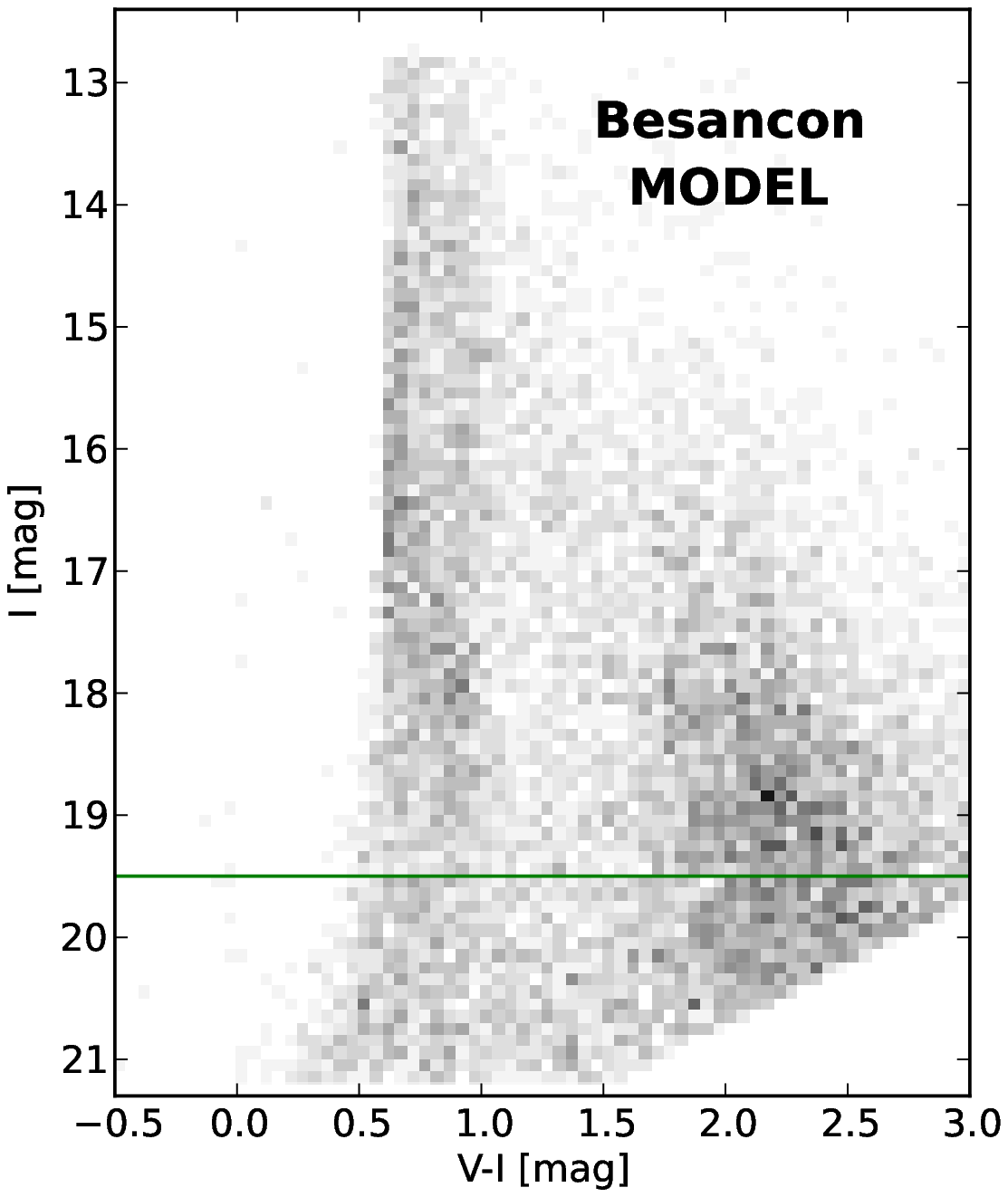} &
\includegraphics[height=4.9cm]{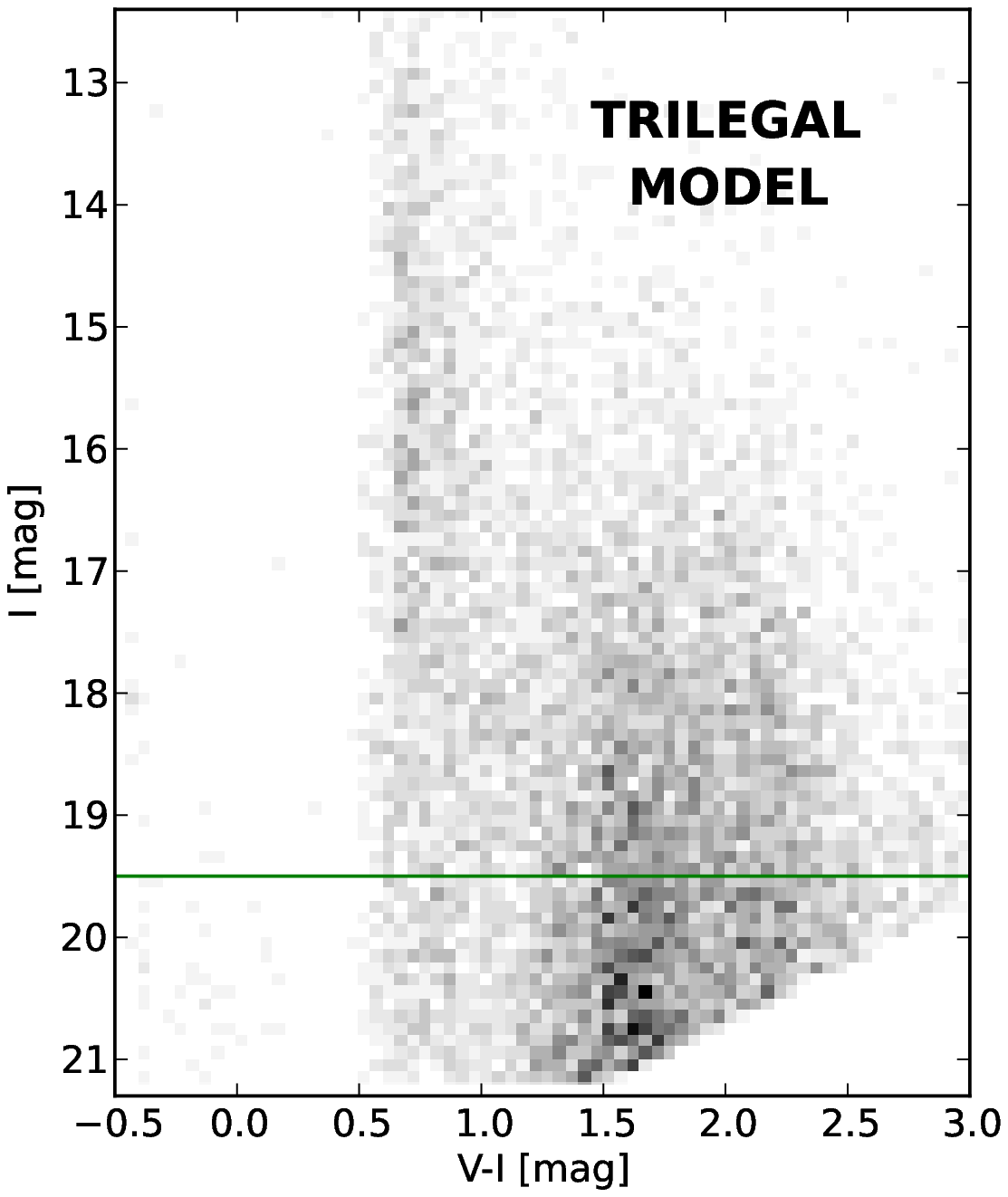} &
\includegraphics[height=4.9cm]{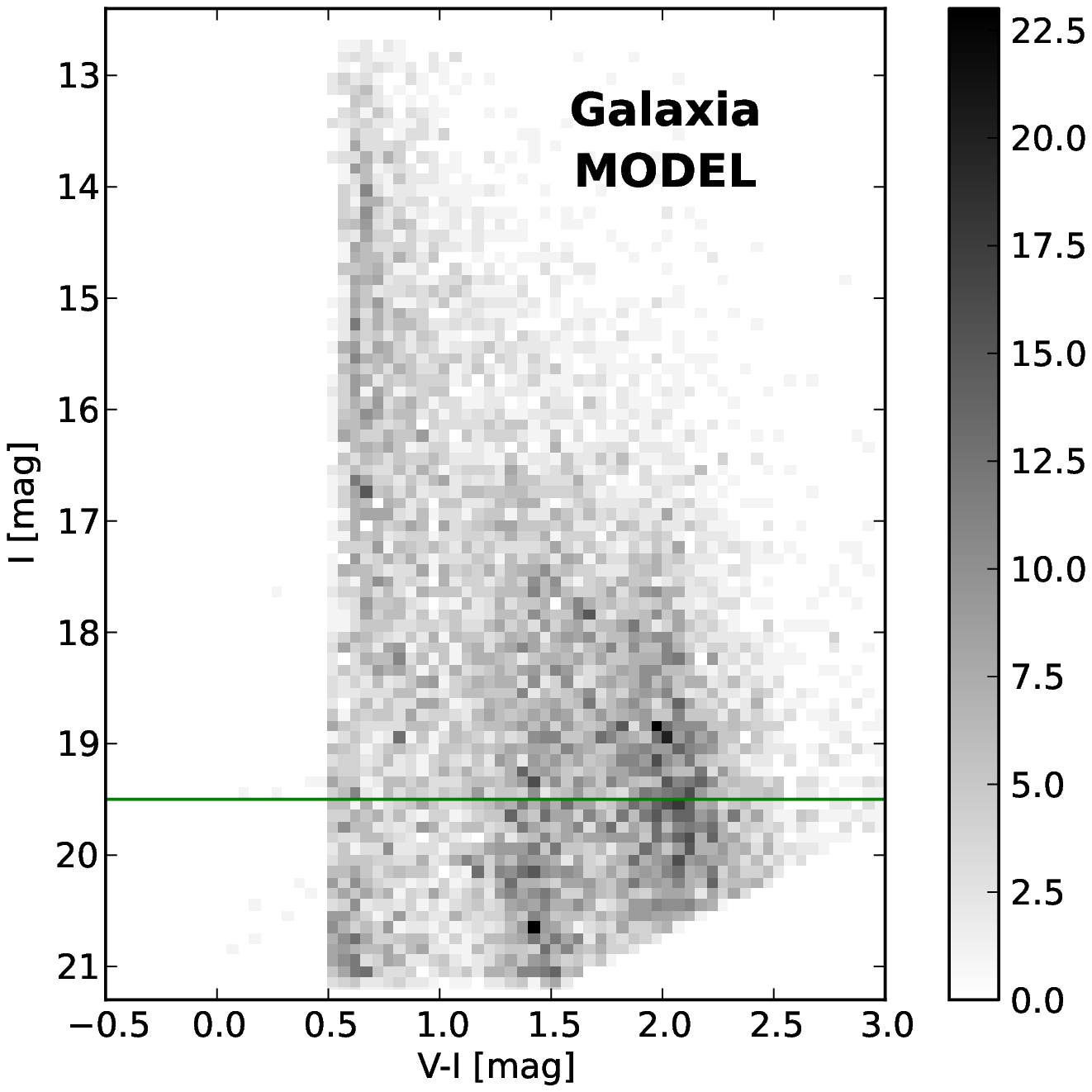} \\
\end{tabular}
\caption{Hess diagrams of LMC694 -- one of the OGLE-IV Galactic fields.
The left panel shows real OGLE-IV data, while the middle and right panels show
Galaxy simulations of the same area on the sky, based on Besan\c{c}on, TRILEGAL
and Galaxia models, respectively. Bin sizes are 0.1 mag in $I$ and 0.05 mag in
$V-I$. The color of each bin corresponds to the number of stars in that bin,
as indicated on the color bar on the right. The horizontal line at
$I=19.5$~mag is drawn to focus attention on the upper part of the diagrams,
which should be almost unaffected by noise.}
\vspace{0.3cm}
\label{fig:gal}
\end{figure*}
\setlength{\tabcolsep}{6pt}
\notetoeditor{(fig:gal) black and white for print, in color only in 
the electronic version}

Next we construct a luminosity function for each field to find data
completeness limits. In Figure~\ref{fig:lumfunct}, we plot luminosity functions
of all 47 MBR fields (both panels). The top panel shows data before the
cleaning process described above and the bottom panel shows data after the
cleaning. We see that the luminosity function shape changed for $19<I<21$.
A small bump around 20.5~mag (top panel), which was mainly due to spurious
detections around bright stars, has almost disappeared,
although not completely, as the uncertainty whether an object is real or not,
grows quickly at faint magnitudes. The bump at $I=18.2$~mag is the red clump
present in some MBR fields, and the bump at $I=22.5$~mag is mostly
due to cosmic rays that had not been removed by the pipeline.
We choose $I$ band magnitudes of 12.8 and 21.2 as the bright and faint
completeness limits for our sample, keeping in mind that the faint end of the
luminosity function is contaminated by spurious detections.

As can be seen in Figure~\ref{fig:fields}, a number of OGLE-IV fields overlap
resulting in multiple detections in those regions. The detection is considered
multiple if the distance between objects is $\leq0\farcs52$ (equivalent of 2
OGLE pixels).
We remove those duplicate objects in a way that we keep the one that has more
epochs, or, if the number of epochs is similar (within $10\%$), has the lower
light curve errors.

The Magellanic Bridge region contains numerous globular and open clusters.
We use the most recent catalog of the Magellanic System clusters
\citep{Bica2008} as a reference to identify all clusters lying within OGLE-IV
fields (189 clusters) and we remove them from the final sample using the mean
value of both dimensions as a cluster diameter.

After all cleaning steps described above, the database is reduced from 12 to
5 million sources (from 5 to 2.5 million sources for stars brighter than
$I=20$~mag).

\subsection{Extinction Correction}

Extinction toward the Magellanic Bridge is generally small. Dust extinction
maps from \cite{Schlegel1998} give \mbox{$E(B-V)$} values in the range
0.02--0.15~mag with the mean value of 0.06~mag. This translates to
\mbox{$E(V-I)$} between 0.03--0.19~mag and the mean of 0.08~mag. Galactic
foreground fields were chosen to have very low color excesses, with
\mbox{$E(V-I)$} not exceeding 0.04~mag. We correct both the MBR and the
calibration fields for extinction using \cite{Schlegel1998} extinction tables
obtained from the NASA/IPAC Infrared Science 
Archive\footnote{http://irsa.ipac.caltech.edu/applications/DUST/}
with the spatial resolution of 0.05~deg.

\subsection{Galactic Foreground Removal}

The Galactic foreground contribution can be accounted for by subtracting
a Galactic Hess diagram\footnote{A Hess diagram is a CMD that has been binned
both in magnitude and color, and the value of each bin is a number of stars
that fell into that bin.} from a Hess diagram of the science field. This
requires either having observations of a purely Galactic field that correspond
to the science data in terms of location, magnitude and color range, and
completeness (ideally from the same telescope), or creating a Galaxy model with
adequate parameters for an area that needs to be cleaned, e.g. the widely used
{\em Besan\c{c}on Model of stellar population synthesis of the Galaxy}
\citep{Robin2003}, or {\em TRILEGAL} \citep{Girardi2005}, or a new promising
tool {\em Galaxia} \citep{Sharma2011} that combines the advantages of the
Besan\c{c}on and TRILEGAL Galaxy models.

On the other hand, there are a number of fields within OGLE-IV sky coverage
suitable as a Galactic foreground representation.
We chose the ones marked with thick lines in Figure~\ref{fig:fields}.
They constitute three groups in terms of Galactic
latitude: six far LMC fields (LMC692-LMC694, LMC701-LMC703) at
$b\approx-36^{\circ}$, four MBR fields (MBR205, MBR206, MBR211, MBR212) at
$b\approx-47^{\circ}$ and four MBR fields (MBR156, MBR157, MBR181, MBR182) at
$b\approx-52^{\circ}$. These fields lie far from both galaxies
and any known dust regions, and their Galactic latitudes cover the latitude
range of the investigated region (all coordinates are listed in
Table~\ref{tab:fields}), so they should be a good representation of the Galactic
population. The left panel in Figure~\ref{fig:gal} shows a Hess diagram of one
of those fields (LMC694). For comparison, the middle and right panels show Hess
diagrams of three Galaxy models generated for the same area that is covered by
field LMC694, using standard model parameters of Besan\c{c}on, TRILEGAL, and
Galaxia Galaxy models (downloaded from  http://model.obs-besancon.fr/,
                      http://galaxia.sourceforge.net/
                 and  http://stev.oapd.inaf.it/cgi-bin/trilegal), respectively.
As expected, data-based CMD is a good representation of the Galactic
population and does not show any features characteristic of the nearby
galaxies (e.g., the main sequence or the red giant branch populations), hence
we will further use it for the Galactic foreground removal, instead of Galaxy
models presented above.

\begin{figure}[tb]
\centering
\includegraphics[height=6.6cm]{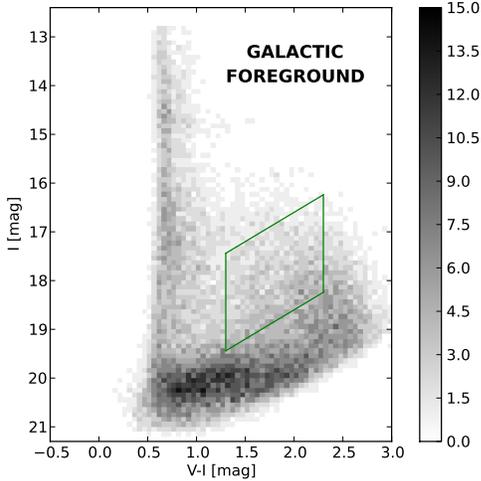}
\caption{Average combined Hess diagrams of the six Galactic OGLE-IV fields
(LMC692-LMC694 and LMC701-LMC703). Bin sizes are 0.1 mag in $I$ and 0.05 mag
in $V-I$. The color of each bin corresponds to the number of stars in that bin,
as indicated on the color bar on the right. Green rhomboid marks the region
used for scaling the diagram before it is subtracted.}
\vspace{0.3cm}
\label{fig:gal_combined}
\end{figure}
\notetoeditor{(fig:gal_combined) black and white for print, in color only in      
the electronic version}

We create OGLE data-based Galactic Hess diagrams by average-combining six/four
fields within each of the three Galactic foreground groups (in order to reduce
pixel noise) and use those as a Galactic foreground representation (see
Figure~\ref{fig:gal_combined}). For each science Hess diagram, we choose
a foreground diagram that is closest to it in terms of $b$ and use it for
Galactic foreground subtraction. When subtracting, we scale the Galactic Hess
diagram such that it contains the same number of stars as the science diagram,
in an area that is expected to consist of the Galactic population only.
We choose this CMD region to be enclosed with a set of lines:
$V-I > 1.3$; $V-I<2.3$; $I<-1.2\times(V-I)+21$; $I>-1.2\times(V-I)+19$
(green rhomboid in Figure~\ref{fig:gal_combined}).
This should account for incompleteness in the observational data.

As a side test, we also subtracted Besan\c{c}on, TRILEGAL and Galaxia Hess
diagrams from science Hess diagrams for a number of MBR fields, using Galaxy
models generated at those exact locations. We then compared the $\chi^2$ of all
subtractions (data, Besan\c{c}on, TRILEGAL, and Galaxia) in the CMD region
occupied by the Galactic population. We found the $\chi^2$ values to be about
twice as large for the models subtractions as compared to the data subtractions.
This additionally supports the use of the far LMC and MBR fields as a 
data-based Galactic foreground representation, instead of a model-based one.

\begin{figure*}[tbhp]
\centering
\includegraphics[width=18cm]{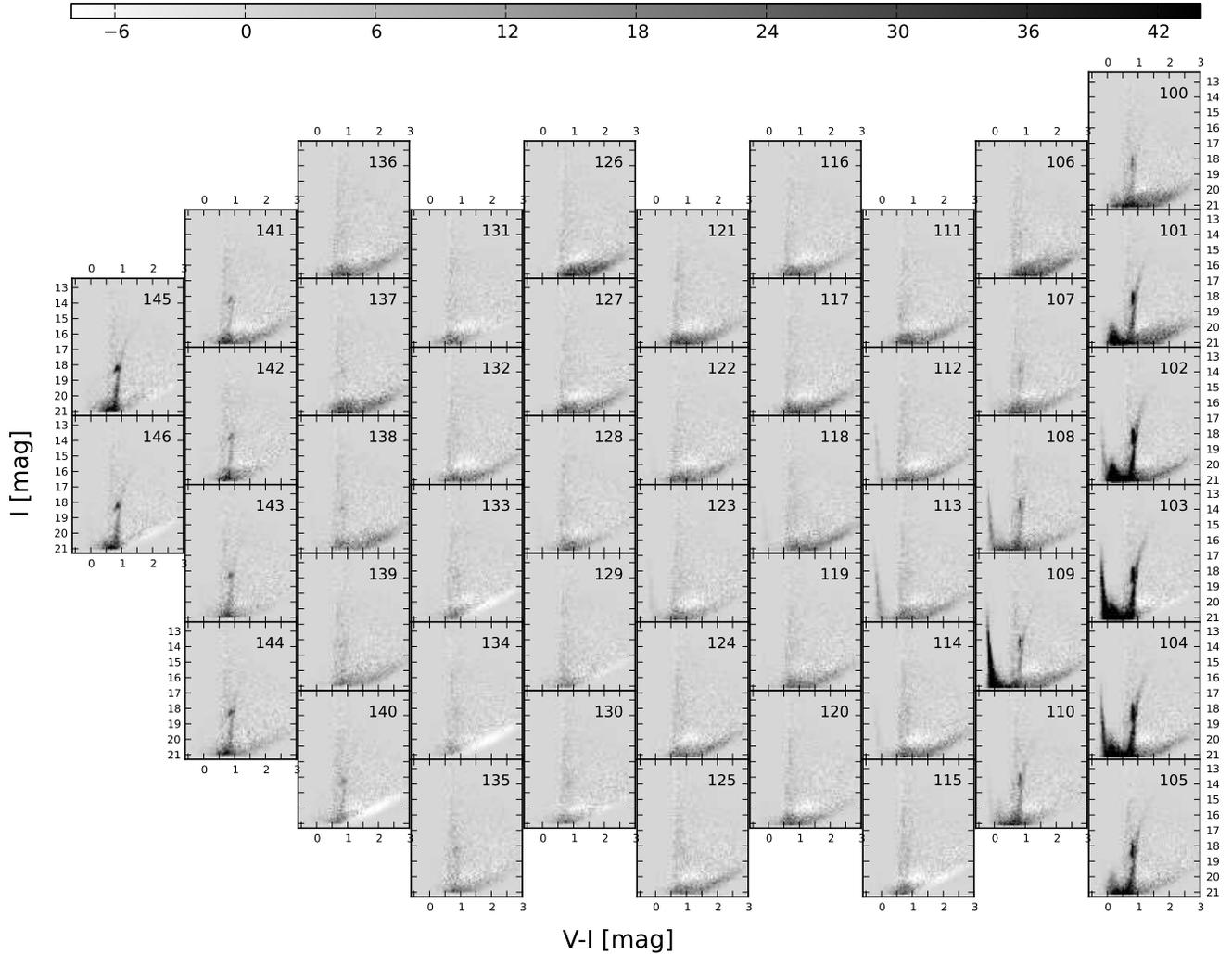}
\caption{Hess diagrams of the main part of the Magellanic Bridge region, after
removing the Galactic foreground contribution. Figure panels are arranged to
reflect OGLE-IV field location in the sky, such that the LMC is to the left and
the SMC is to the right (compare with Figure~\ref{fig:fields}). Each panel
presents a CMD of the entire OGLE-IV field. Bin sizes are 0.1 mag in $I$ and
0.05 mag in $V-I$. The color of each bin corresponds to the number of stars
in that bin, as indicated on the color bar at the top of the figure.}
\label{fig:cmds}
\vspace{0.2cm}
\end{figure*}

\section{Results and Discussion}

\subsection{Color--Magnitude Diagrams}

Figure~\ref{fig:cmds} shows Galaxy-subtracted Hess diagrams of 47 fields in the
main part of the Bridge region (fields MBR100 -- MBR146). Figure panels are
arranged in a way that reflects each field's location on the sky (compare with
Figure~\ref{fig:fields}). For the majority of presented fields there is
a region of over-subtraction at $I\sim20$~mag and $V-I$ between $1-2$~mag,
and a region of under-subtraction below it, which are caused by a lower
magnitude limit of the Galactic fields as compared to the MBR fields (due to
a smaller number of observations).

The young population (YP, $V-I\lesssim0$) is very prominent on the SMC side and
fades as we look further into the MBR, reaching about half way in the central
part (field MBR123). This is consistent with previous findings
(e.g. \citealt{Harris2007}), but the exact extent and density distribution of
the YP across the Bridge have not been known so far. Older and intermediate-age
populations, i.e. red giant branch (RGB) and red clump (RC) stars,
respectively, are present on both the LMC and the SMC side, and what we see
is most probably the extent of these galaxies into the MBR. Interestingly, the
shape of the RC distribution on the CMD is round on the LMC side and vertically
elongated on the SMC side. This elongation is caused by a large line-of-sight
depth, rather than a presence of blue loops stars, as shown by
\cite{Nidever2013}.

We will discuss the population distributions in greater detail in the
following sections.

\subsection{Population Selection Regions}

\setlength{\tabcolsep}{0pt}
\begin{figure}[tb]
\centering
\includegraphics[height=8cm]{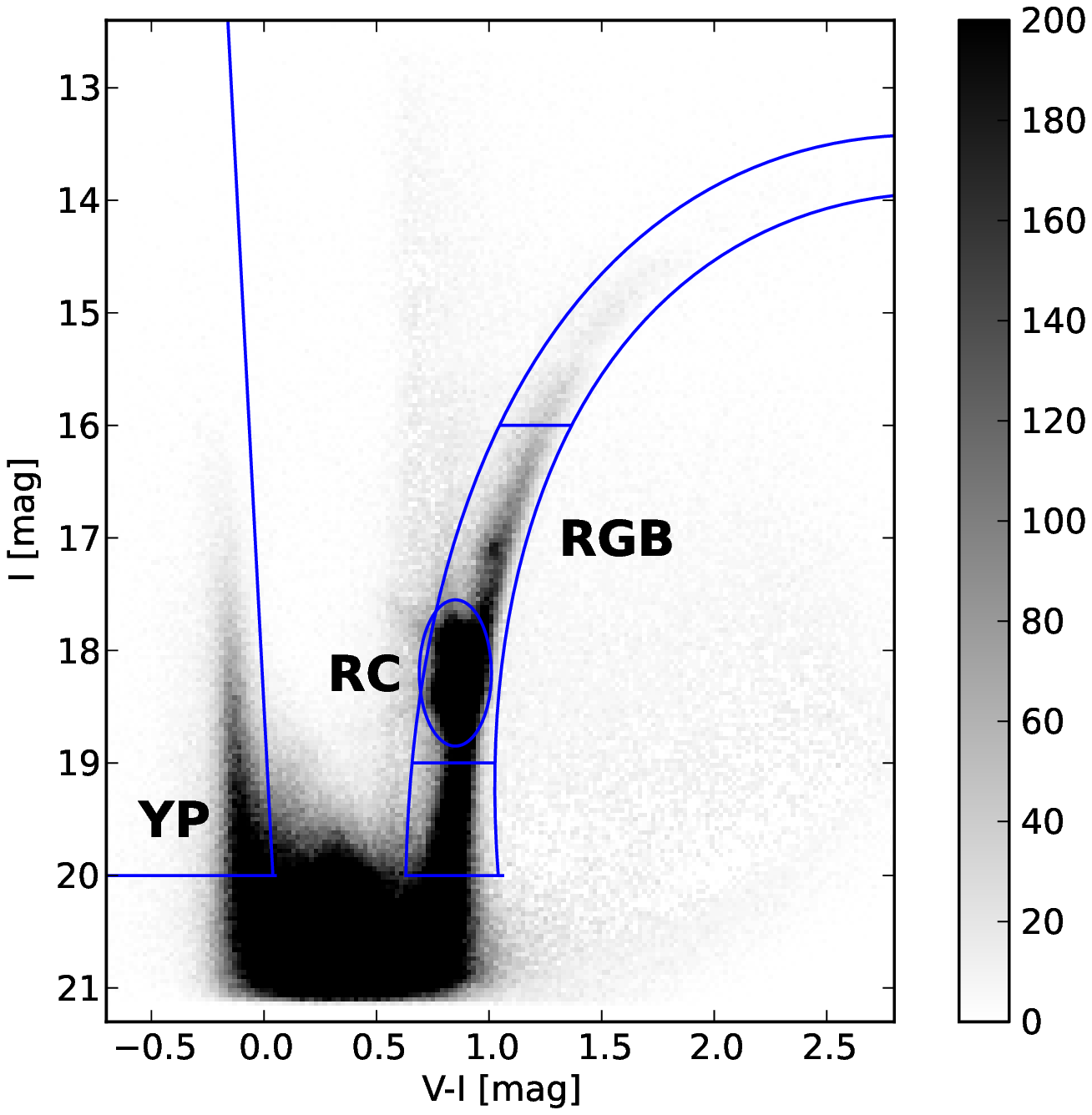} 
\caption{Cumulative Galaxy-subtracted Hess diagrams of a number of fields in
the MBR. Lines mark regions occupied by the young population (YP), RC and RGB
stars. Line equations are: $I \geq 38\times(V-I)+18.5$, and $I\leq20$ for
the YP; $(V-I-0.85)^2/0.16^2 + (I-18.2)^2/0.65^2 \leq 1$ for the RC;
$(V-I-4.5)^2/3.5^2 + (I-24.9)^2/11.6^2 \leq 1$,
$(V-I-4.5)^2/3.1^2 + (I-23.1)^2/9.3^2 \geq 1$, and $V-I\leq3.0$, and $I\leq20$
for the RGB stars. RGB stars are further subdivided into top, middle and bottom
part at $I=19$ and $I=16$~mag. To increase resolution binning has been reduced
to 0.04~mag in $I$ and 0.02~mag in $V-I$. As in previous figures, the color of
each bin corresponds to the number of stars in that bin, as indicated on the
color bar on the right.}
\label{fig:regions}
\vspace{0.3cm}
\end{figure}
\setlength{\tabcolsep}{6pt}
\notetoeditor{(fig:regions) black and white for print, in color only in      
the electronic version}

\begin{figure}[tbh]
\centering
\includegraphics[width=8.5cm]{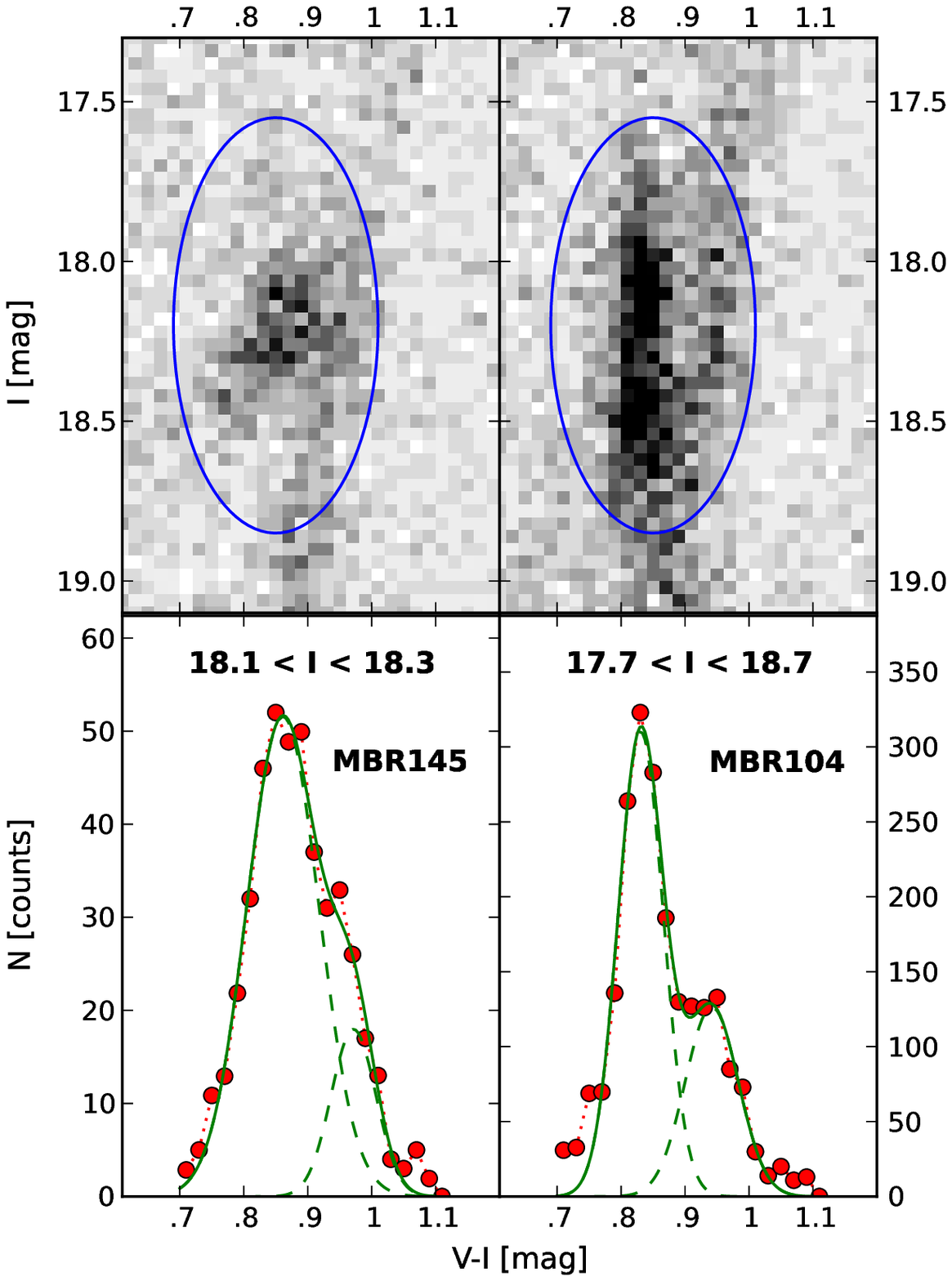}
\caption{A comparison of elongated and compact RC for two fields on the
opposite sides of the Bridge: MBR104 (left, close to the LMC) and MBR145
(right, close to the SMC). To increase resolution binning has been reduced
to 0.04~mag in $I$ and 0.02~mag in $V-I$.
Top panels show Galaxy-subtracted Hess diagrams of the RC region.
Bottom panels show $V-I$ histograms for the same fields, where the data have
been summed along the magnitude axis within the range indicated on the images.
Green solid line shows a two Gaussian fit to the data (red dots), while green
dashed lines show each of the two Gaussian distributions.
}
\label{fig:regions_rc_rgb}
\end{figure}
\notetoeditor{(fig:regions_rc_rgb) black and white for print, in color only in
the electronic version}

Selection regions for the three main stellar populations (YP, RC, and RGB) are
shown on a Hess diagram in Figure~\ref{fig:regions}, with line equations in the
figure caption. To avoid contamination from stars belonging to the main sequence
of the two galaxies, we set a limit $I<20$~mag
for all groups. The RC ellipse is strongly elongated to include all RC stars
in fields that have a large line-of-sight depth. As shown by \cite{Nidever2013}
for the SMC periphery, this is strictly a distance effect and does not indicate
a presence of blue loop stars. A region occupied by RGB stars was further
subdivided into top ($I<16$~mag), middle ($16<I<19$~mag), and bottom ($I>19$~mag)
parts to separate asymptotic giant branch (AGB) and younger RGB stars (top RGB
region) stars from old, low-mass stars (bottom RGB region).

Areas occupied by the RC and the RGB stars overlap significantly and this has
to be taken into account when extracting RC and middle RGB groups from the
sample. In order to separate these populations we divide the RC ellipse into
two regions in $V-I$. Stars bluer than the separation value are included into
the RC population, while stars redder than the separation value are included
into the RGB population. Top panels of Figure~\ref{fig:regions_rc_rgb} show
close-ups of the RC region in the Hess diagrams of two exemplary fields lying
on the opposite sides of the Bridge (MBR104 and MBR145), one having an
elongated RC and the other a compact RC. Bottom panels show $V-I$ histograms
of those close-ups, where for each $V-I$ value data were summed over a range of
magnitudes covering the RC (indicated on the images). As an attempt to
determine a separation value mentioned above, we fit a function composed of
two Gaussian distributions to the summed histograms. The fits are plotted
with green lines in Figure~\ref{fig:regions_rc_rgb}.
In the case of field MBR104 (elongated RC), the fit yields:
$\mu_{(V-I),RC} = 0.83\ \rm{mag}$, $\sigma_{(V-I),RC} = 0.04\ \rm{mag}$, and
$\mu_{(V-I),RGB} = 0.94\ \rm{mag}$, $\sigma_{(V-I),RGB} = 0.04\ \rm{mag}$,
and the Gaussian cross at $V-I\approx 0.89$~mag.
In the case of field MBR145 (compact RC), the fit results are:
$\mu_{(V-I),RC} = 0.86\ \rm{mag}$, $\sigma_{(V-I),RC} = 0.06\ \rm{mag}$, and
$\mu_{(V-I),RGB} = 0.97\ \rm{mag}$, $\sigma_{(V-I),RGB} = 0.04\ \rm{mag}$,
and the Gaussian cross at $V-I\approx 0.94$~mag.
We adopt the crossing point of the two functions as a separation value
between the RC and the RGB, within the ellipse surrounding the RC region.
Outside this ellipse, we use the regions as shown in Figure~\ref{fig:regions}.

Using a single crossing point as a separation between the RC and the RGB is
a rough estimation that does not take into account the fact that
the overlap fraction of the RC and the RGB changes when moving along the $I$
axis, due to an inclination of the RGB. However, this value is averaged
over the magnitude range covering the whole overlap, which is larger than
average at fainter magnitudes and smaller at brighter magnitudes. So if
the densities within each population do not change very differently with
magnitude over this magnitude range, an average value should be
a sufficient approximation.

\subsection{Two-dimensional Density Maps}

In the last step before we construct density maps, we subdivide all data into
smaller regions (in right ascension and declination) to increase the spatial
resolution of the maps. We chose a 0.335~deg$^2$ square subfield (instead of
1.4~deg$^2$ OGLE-IV field) as a compromise between better spatial resolution
and good count statistics of individual Hess diagrams.

Figures~\ref{fig:maps_yp}, \ref{fig:maps_rc}, \ref{fig:maps_rgb_top}, 
and~\ref{fig:maps_rgb_bot} show spatial density maps of the YP, RC and RGB
stars in the region covering the whole MBR and two-field wide stripes of the
LMC and the SMC adjacent to the MBR. All maps are drawn using a Hammer
equal-area projection\footnote{http://en.wikipedia.org/wiki/Hammer\_projection}
centered at $\alpha=3.3$~h and $\delta=-70$~deg. The color-coded value of each
``pixel'' is a logarithm of the number of stars per square degree area, while
each ``pixel'' area is approximately $0.335\deg^2$.
This value has been corrected for the OGLE data completeness factor, which
originates from gaps between OGLE-IV fields, as well as from horizontal and
vertical gaps between the 32 chips in the OGLE-IV camera. The completeness
factor also takes into account masked regions around bright stars and stellar
clusters. It varies between 80\% and 98\% for a typical subfield, with a mean
value of 93\%, but can be as low as a few percent, if the subfield happened to
fall close to one of the larger gaps between fields.

The maps also show an approximate location of the LMC disk and the main stellar
body of the SMC, marked with white ellipses at
$\alpha=05^{\rm h}29^{\rm m}$, $\delta=-69^{\circ}30'$ (LMC), and 
$\alpha=00^{\rm h}54^{\rm m}$, $\delta=-72^{\circ}57'$ (SMC).
The white cross marks the SMC center of the outer SMC population found by
\cite{Nidever2011} at $\alpha=01^{\rm h}00^{\rm m}31^{\rm s}$ and
$\delta=-72^{\circ}43'11''$.

A full list of all number densities for the YP, RC and RGB stars together with
their coordinates is available on-line from the OGLE website
http://ogle.astrouw.edu.pl, and a few exemplary lines are listed in
Table~\ref{tab:densities}.

\newpage
\subsubsection{The Young Population}

Figure~\ref{fig:maps_yp} shows a density map of the young ($\lesssim1$~Gyr)
population, where only $\geq 2\sigma$ detections have been plotted. Color
contours mark a number of neutral hydrogen (\HI) emission levels. The
\HI~emission was integrated over the velocity range 80 < v < 400 km/s, where
each contour represents the \HI~column density twice as large as the
neighboring contour. \HI~column densities are in the range
$10^{20}-4\times 10^{21} {\rm cm}^{-2}$. Data were taken from the LAB survey
of Galactic \HI~\citep{Kalberla2005}.

We see that the YP is present mainly in the western part of the main MBR
($\alpha \leq 3$~h), which is consistent with previous findings
(\citealt{Irwin1985}, \citealt{Irwin1990}, \citealt{Demers1998},
\citealt{Harris2007}) and simulations \citep{Besla2012}.
However, we discover that there is a
non-negligible young population in its eastern part, in the direction of
the LMC ($3\lesssim \alpha \lesssim 4.2$~h, $-74\lesssim \delta
\lesssim -73$~deg), at the level of 25-90 stars/deg$^2$, which is
significantly higher than the median background level estimated from 40
most southern fields to be 1 star/deg$^2$ with a standard deviation of 6
stars/deg$^2$, meaning that these are $4-14\sigma$ detections. This is
the first time that the YP is seen in the eastern Bridge, thus showing that
there is a continuous stream of young stars connecting the two galaxies.

It is also worth noting that preliminary search for pulsating stars in
the main MBR area resulted in the discovery of four short period
classical Cepheids distributed along eastern and central parts of
the bridge, which is reassuring. A more detailed analysis of pulsating stars
in the MBR will be published in the forthcoming papers.

We also show details of the YP distribution within the MBR, which was not
possible before, as there was no optical survey that fully mapped the entire
region.
We can clearly distinguish two major overdensities in the western MBR, one
closer to the SMC at $\alpha \approx 2$~h and $\delta \approx-74$~deg, and the
other, approximately mid-way between the Clouds at $\alpha \approx 2.9$~h and
$\delta \approx-73.5$~deg. We call the latter an ``OGLE island'' since it is
fairly isolated from the main YP part in the MBR. Both overdensities correspond
well to the yellow and green \HI~contours, and the shape and density
distribution of the OGLE island is surprisingly well matched by the shape of
the overplotted \HI~density contour, showing that star formation was indeed
most effective in areas of high \HI~surface densities. Interestingly, the gap
between the OGLE island and the main part of the YP is not reflected by
accordingly lower \HI~surface density.

\setlength{\tabcolsep}{5pt}
\begin{deluxetable*}{ccccrrrrr}
\tablecaption{\footnotesize Number densities for three stellar populations:
the young population (YP), the red clump (RC), and the red giant branch (RGB)
stars. \label{tab:densities}}\\ 
\tablehead{ 
\colhead{$\alpha$} & \colhead{$\delta$} & \colhead{$l$} & \colhead{$b$} &
\colhead{YP} & \colhead{RC} & \colhead{RGB top} & \colhead{RGB mid} & \colhead{RGB bottom} \\
\colhead{[h]} & \colhead{[deg]} & \colhead{[deg]} & \colhead{[deg]} &
\colhead{[stars/deg$^2$]} & \colhead{[stars/deg$^2$]} & \colhead{[stars/deg$^2$]} &
\colhead{[stars/deg$^2$]} & \colhead{[stars/deg$^2$]}
}
\startdata
1.3007 & $-77.937$ & 301.145 & $-39.099$ &    26 &   279 &    50 &   536 &   418\\
1.3171 & $-76.661$ & 300.848 & $-40.355$ &    45 &   773 &    65 &   851 &   698\\
1.3299 & $-75.385$ & 300.545 & $-41.612$ &   253 &  2448 &   185 &  2530 &  2111\\
1.3398 & $-74.109$ & 300.234 & $-42.867$ &  3902 &  4974 &   299 &  4820 &  3913\\
1.3615 & $-79.988$ & 301.289 & $-37.044$ &     7 &   171 &    48 &   301 &   211\\
... & ... & ... & ... & ... & ... & ... & ... & ...
\enddata
\tablecomments{\footnotesize
Median background levels are $1\pm6$~stars/deg$^2$ for the YP,
$36\pm25$~stars/deg$^2$ for the RC, $16\pm12$~stars/deg$^2$ for the top RGB,
$63\pm45$~stars/deg$^2$ for the middle RGB, and $112\pm65$~stars/deg$^2$ for
the bottom RGB region. A full version of this table, containing number densities
in all 754 subfields in the Magellanic Bridge region is available on-line
from the OGLE website http://ogle.astrouw.edu.pl.}
\end{deluxetable*}
\setlength{\tabcolsep}{6pt}

The eastern part of the young population is not as well matched by the
\HI~density contours as the western part -- it seems that the stars follow the
northern edge of the contours more than the southern, and appear to be a
continuation of the OGLE island toward the LMC along $\delta \approx-73.5$~deg.
\cite{Harris2007} found that the eastward extent of the YP is truncated at
$\alpha \simeq 3$~h, but this can be attributed to his sparse field
distribution in the MBR that could miss areas of higher stellar densities.
On the other hand, \cite{Harris2007} noted that his result is consistent with
the fact that the \HI~surface density in the eastern Bridge falls below the
critical threshold for star formation of about
$5 \times 10^{21} {\rm cm}^{-2}$ \citep{Kennicutt1989}. This would imply either
that the stars we see were not formed in this area or that the \HI~surface
density in the past was in fact sufficient for a star formation episode.

\subsubsection{The Intermediate-Age and Old Populations}

\begin{figure*}[htp]
\centerline{\includegraphics[width=18.5cm]{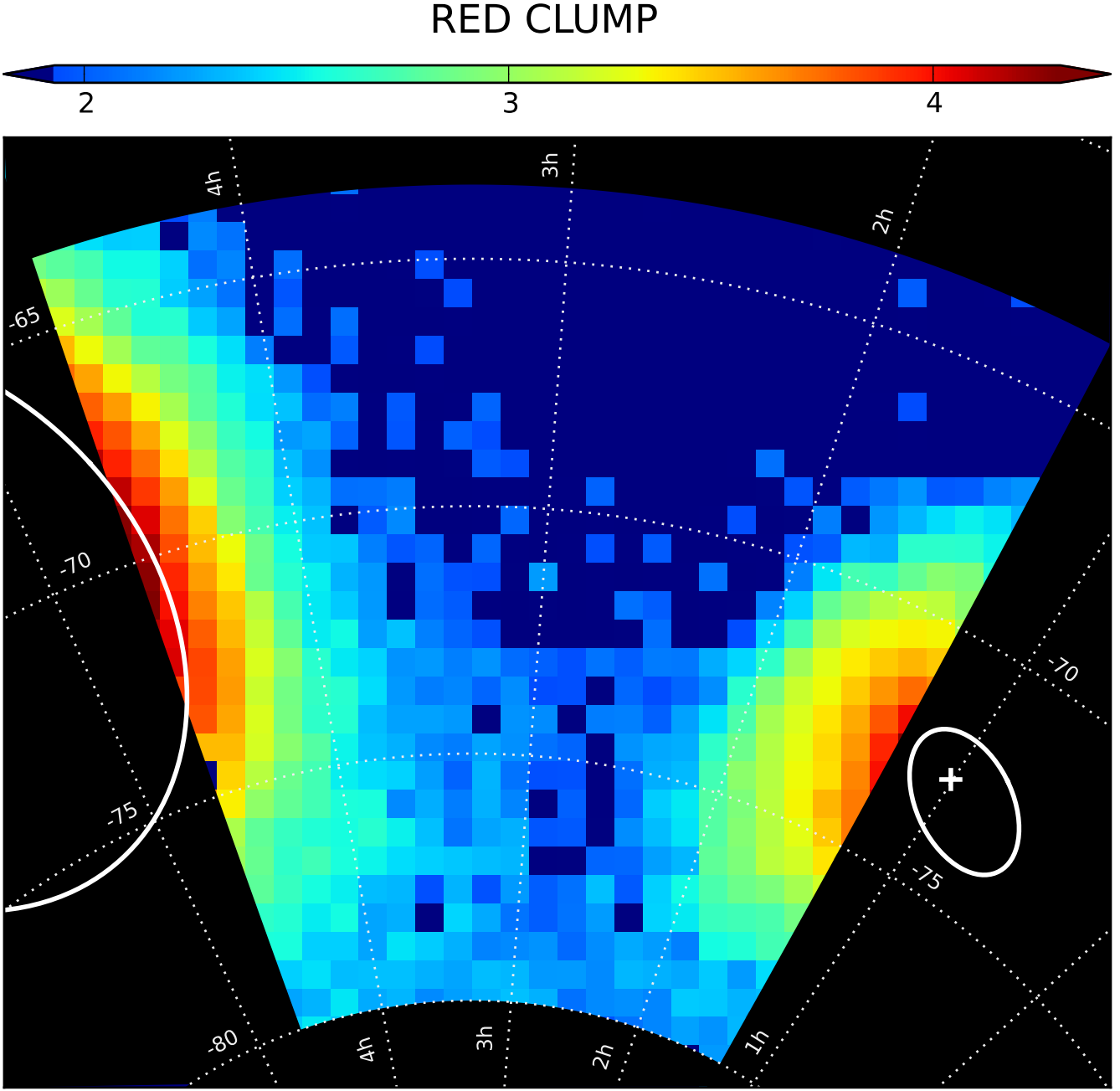}}
\caption{
Spatial density maps of the Red Clump stars in the Magellanic Bridge region,
in a Hammer equal-area projection centered at $\alpha=3.3$~h and
$\delta=-70$~deg. The color-coded value of each ``pixel'' is a logarithm of
the number of stars per square degree area (as indicated on the color bar at
the top of the figure), while each ``pixel'' area is $\sim 0.335\deg^2$.
A median background level was estimated from 50 northern fields to be
36~stars/deg$^2$ with a standard deviation of 25~stars/deg$^2$. Detections
weaker than $2\sigma$ above the median background level have been given the
background color. All number densities are listed in Table~\ref{tab:densities}.
An approximate location of the LMC disk and the main stellar body of the SMC
are marked with white ellipses centered at $\alpha=05^{\rm h}29^{\rm m}$,
$\delta=-69^{\circ}30'$, and $\alpha=00^{\rm h}54^{\rm m}$,
$\delta=-72^{\circ}57'$, respectively.
The white cross marks the SMC center of the outer SMC population found by
\cite{Nidever2011} at $\alpha=01^{\rm h}00^{\rm m}31^{\rm s}$ and
$\delta=-72^{\circ}43'11''$.
}
\label{fig:maps_rc}
\end{figure*}

\begin{figure*}[htp]
\centerline{\includegraphics[width=18.5cm]{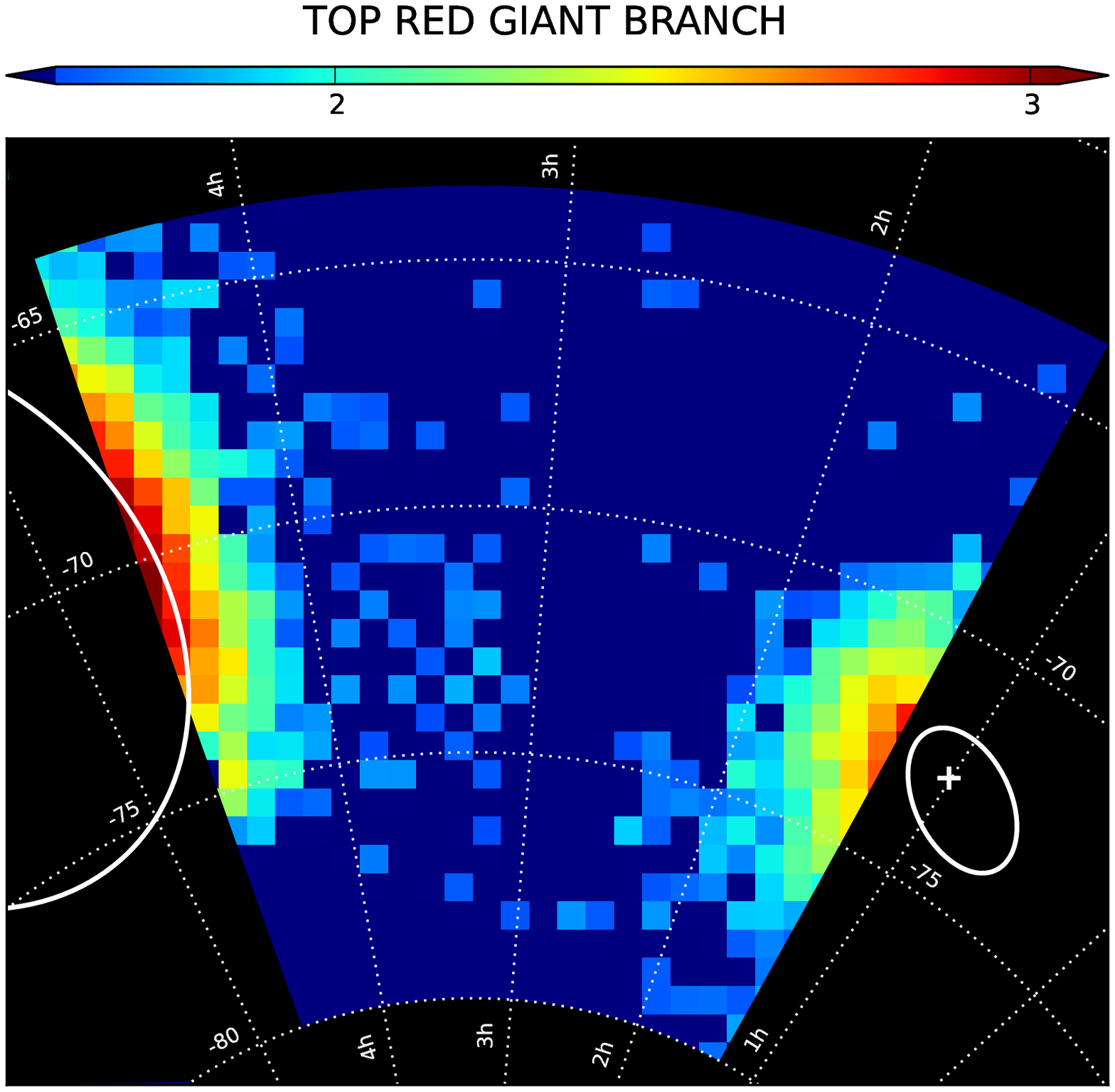}}
\caption{
Spatial density maps of the top part of the Red Giant Branch stars
(I<16~mag, V-I<3~mag) in the Magellanic Bridge
region, in a Hammer equal-area projection centered at $\alpha=3.3$~h and
$\delta=-70$~deg. The color-coded value of each ``pixel'' is a logarithm of
the number of stars per square degree area (as indicated on the color bar at
the top of the figure), while each ``pixel'' area is $\sim 0.335\deg^2$.
A median background level was estimated from 50 northern fields to be
16~stars/deg$^2$ with a standard deviation of 12~stars/deg$^2$. Detections
weaker than $2\sigma$ above the median background level have been given the
background color. All number densities are listed in Table~\ref{tab:densities}.
An approximate location of the LMC disk and the main stellar body of the SMC
are marked with white ellipses centered at $\alpha=05^{\rm h}29^{\rm m}$,
$\delta=-69^{\circ}30'$, and $\alpha=00^{\rm h}54^{\rm m}$,
$\delta=-72^{\circ}57'$, respectively.
The white cross marks the SMC center of the outer SMC population found by
\cite{Nidever2011} at $\alpha=01^{\rm h}00^{\rm m}31^{\rm s}$ and
$\delta=-72^{\circ}43'11''$.
}
\label{fig:maps_rgb_top}
\end{figure*}

Since our CMDs do not reach the main sequence turn-off, we are unable to
resolve age-metallicity degeneracies and estimate ages of observed populations
-- isochrone fitting for fields closer to the MCs is consistent with
populations as old as 13~Gyr and as young as 1~Gyr. In such case we will use
red clump stars (ages $\sim$1 to a few~Gyr), and the top part of the red giant
branch that contains young RGB and AGB stars (ages $\sim500$~Myr to a couple
Gyr), as tracers of an intermediate-age population. 
The bottom part of the RGB contains oldest stars that only started evolving
off the Main Sequence and we will use this group as a tracer of an old
stellar population in the Magellanic System.
Figures~\ref{fig:maps_rc} and~\ref{fig:maps_rgb_top} show density maps of the
RC and the top of the RGB, respectively, that represent an intermediate-age
population. Figure~\ref{fig:maps_rgb_bot} shows a density map of the bottom part
of the RGB that should contain mostly an old population.
All maps show detections $\geq 2\sigma$ only.

\begin{figure*}[htp]
\centering
\vspace{-0.5cm}
\begin{tabular}{cc}
\includegraphics[width=7cm]{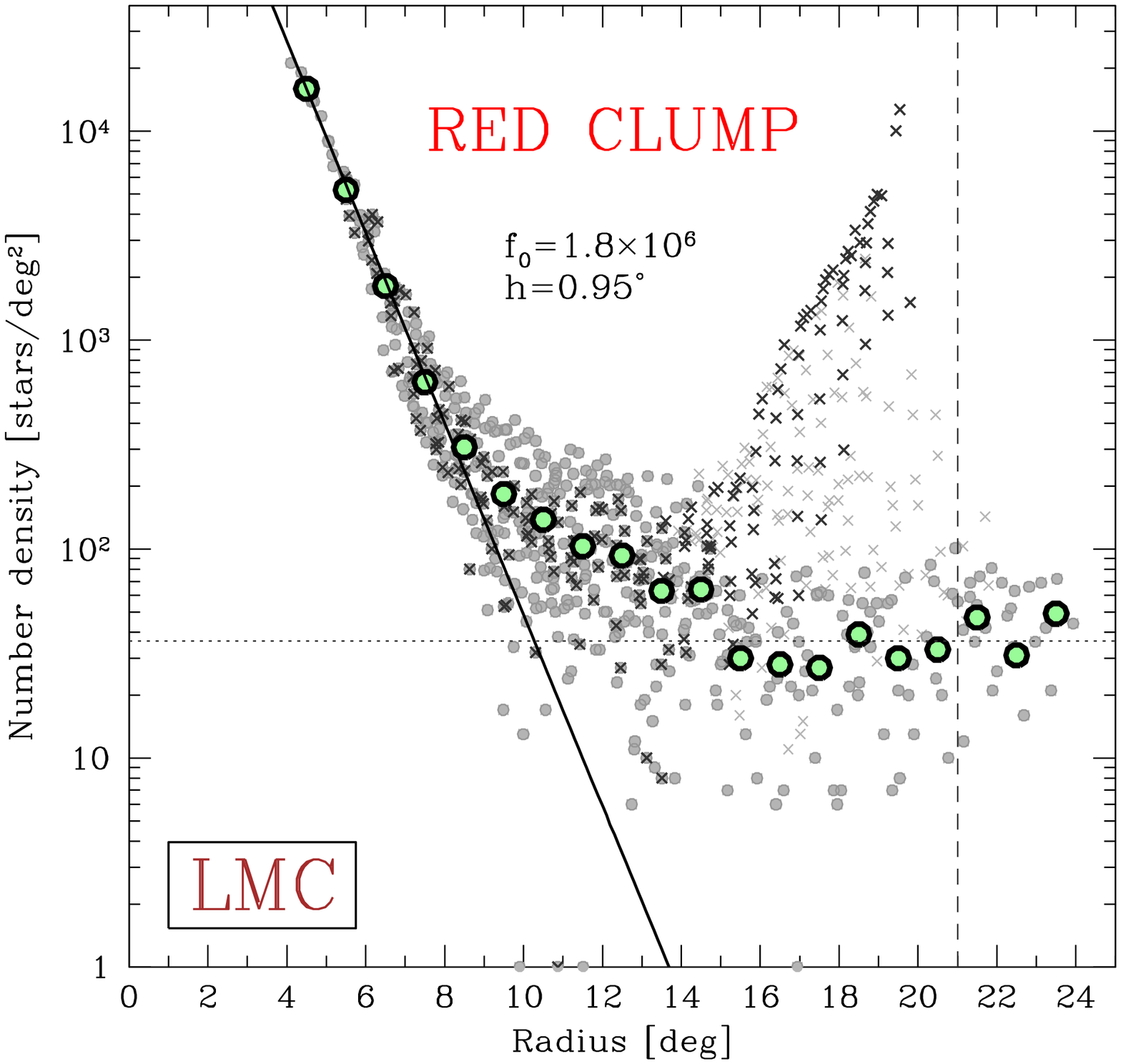} &
\includegraphics[width=7cm]{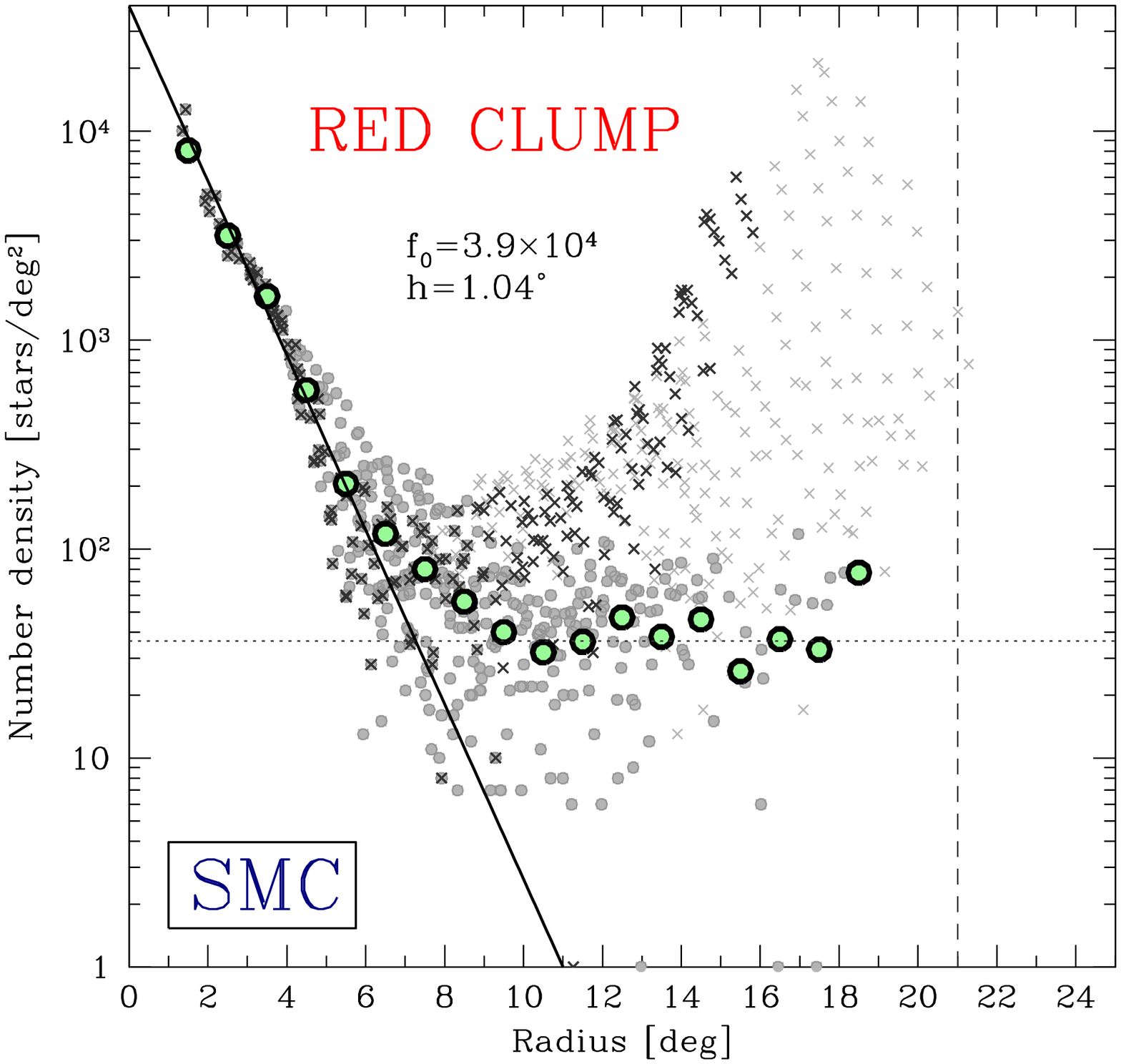} \\ [-4ex]
\includegraphics[width=7cm]{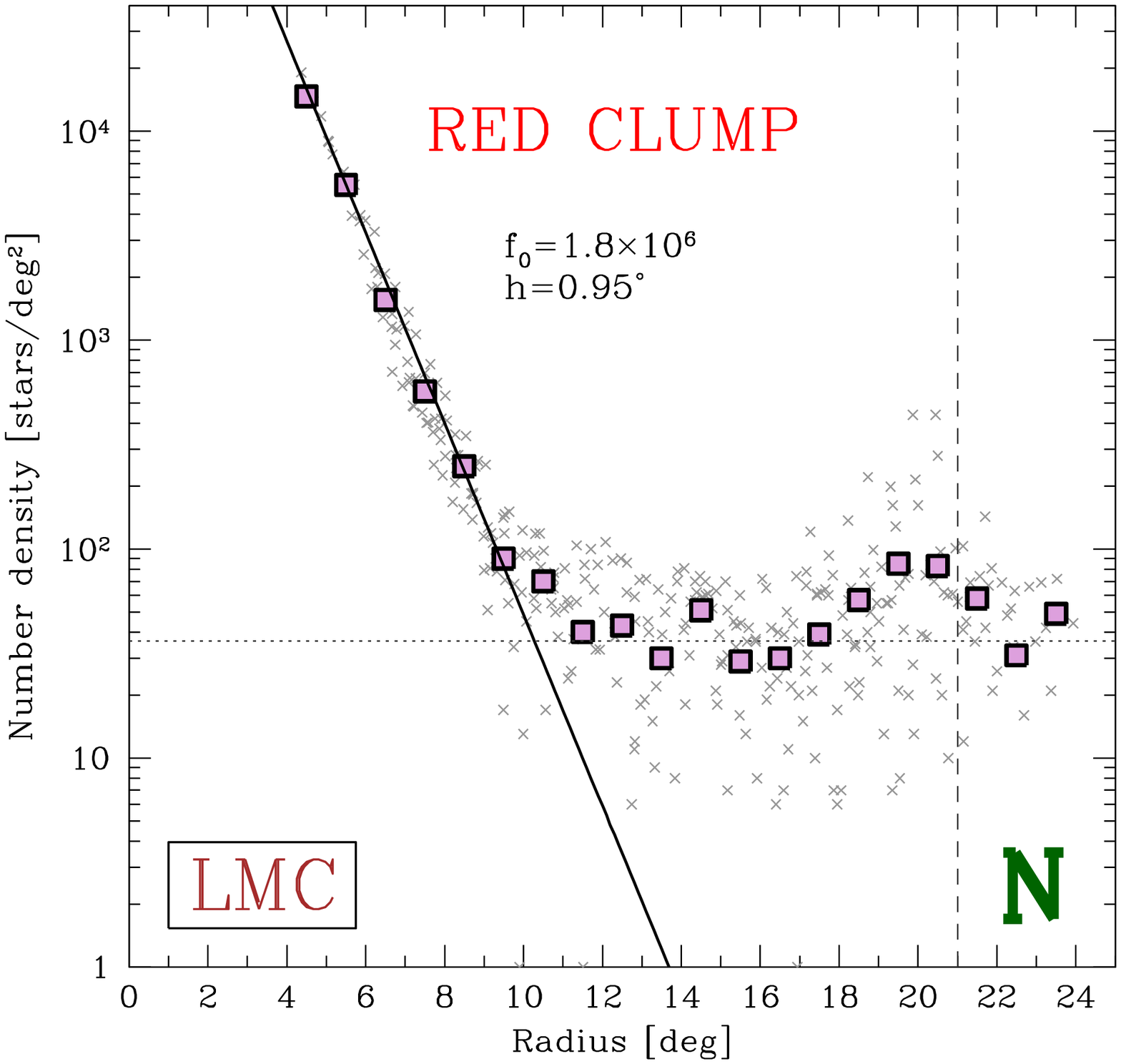} &
\includegraphics[width=7cm]{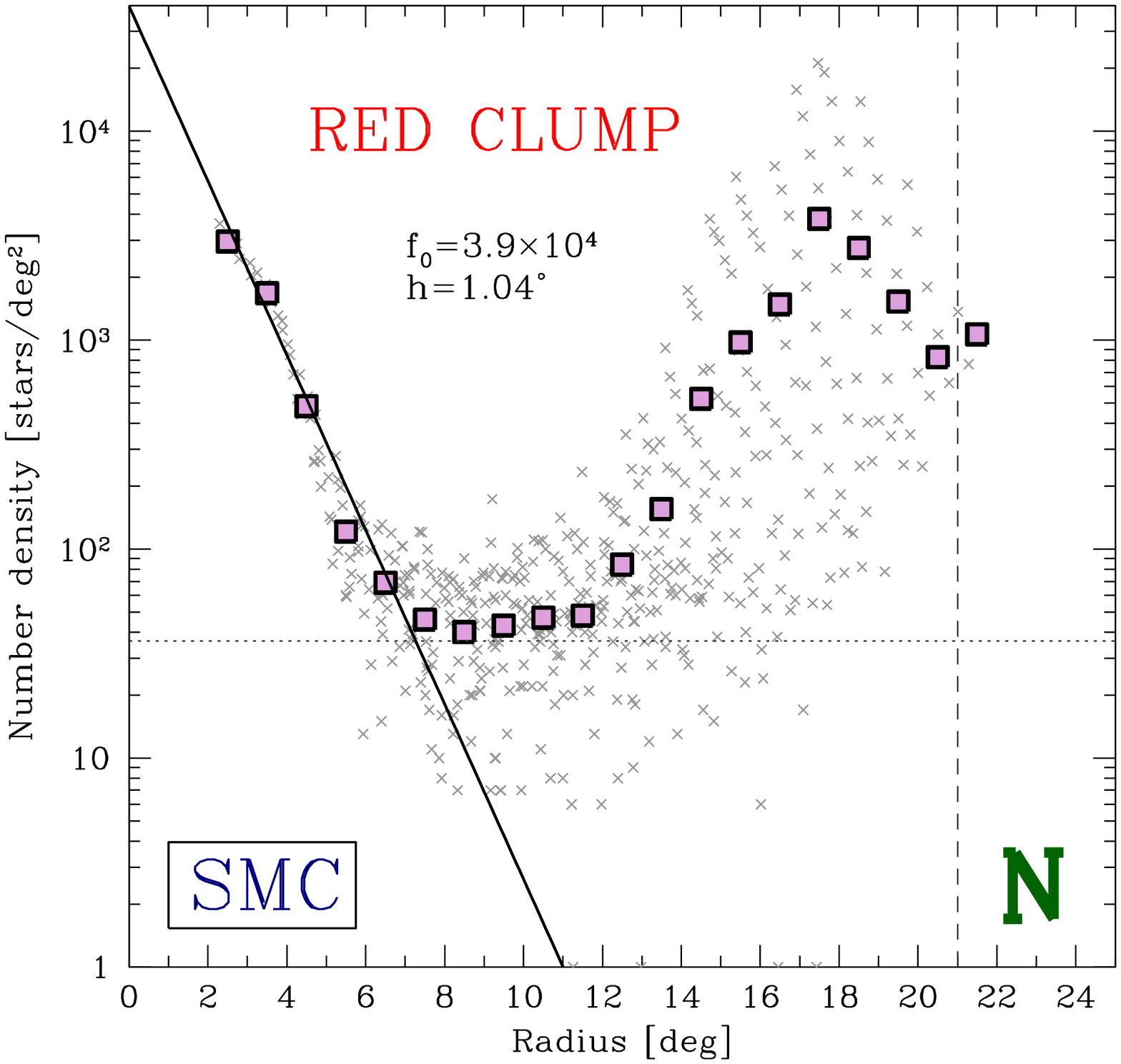} \\ [-4ex]
\includegraphics[width=7cm]{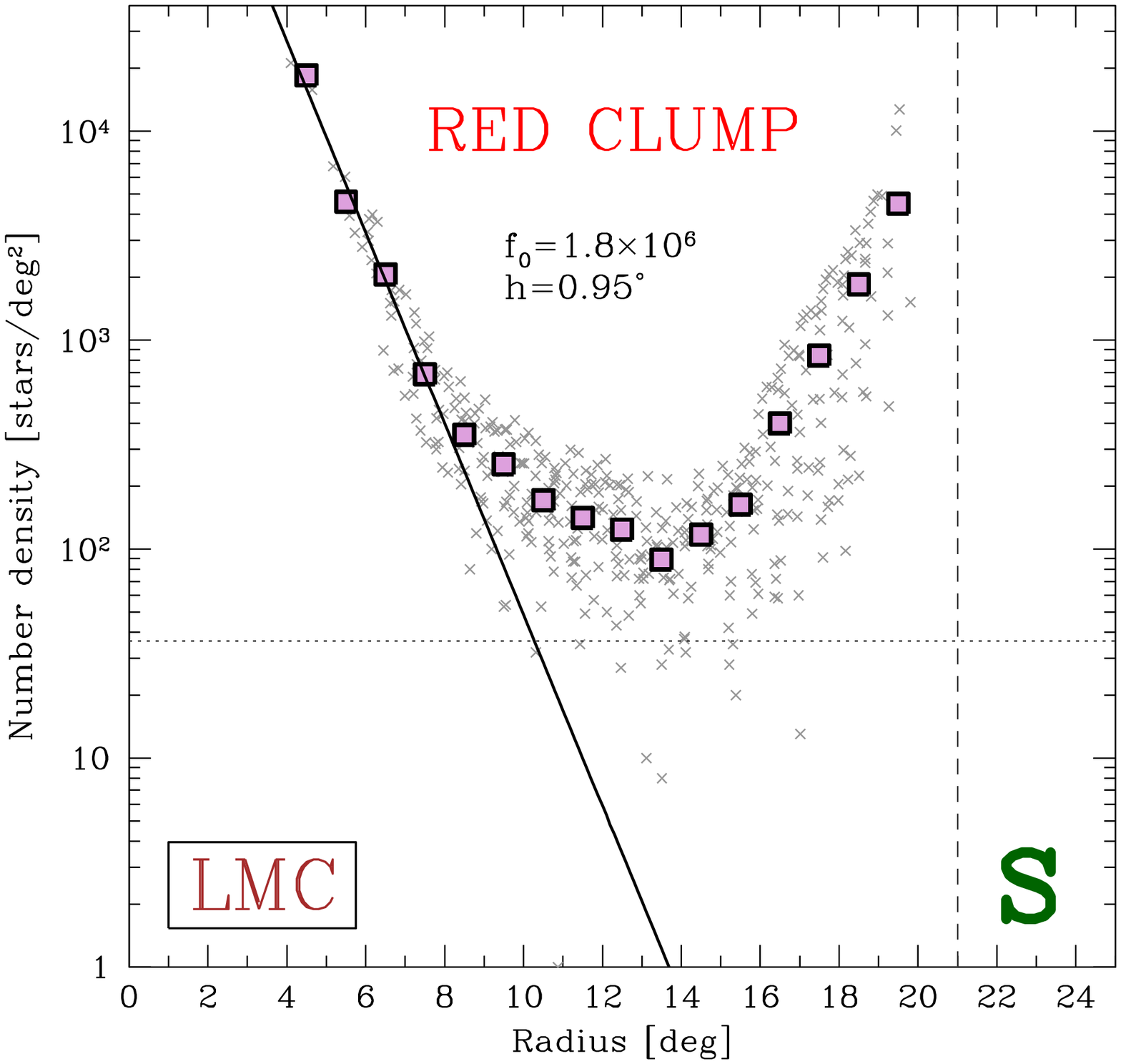} &
\includegraphics[width=7cm]{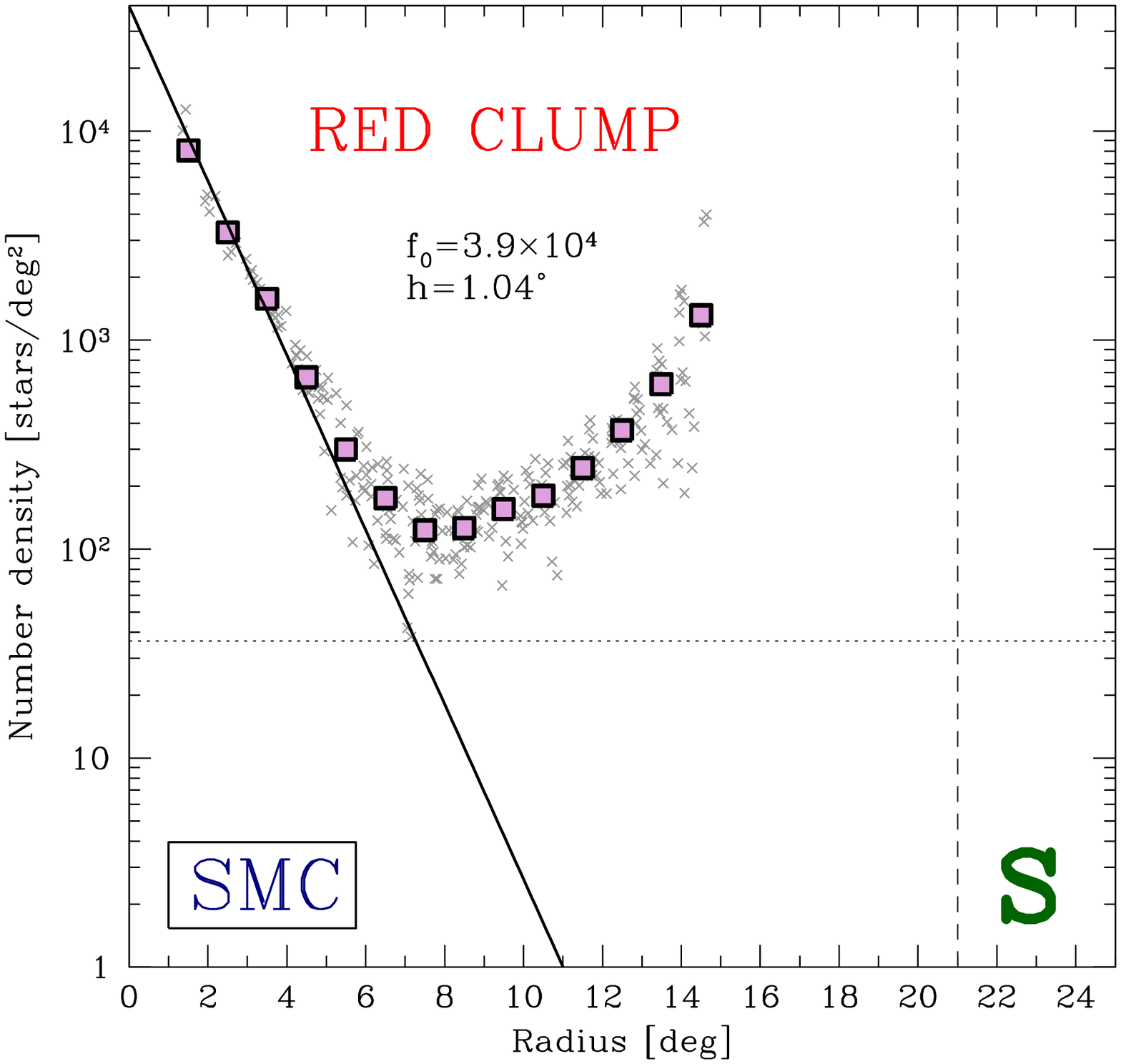} \\
\end{tabular}
\caption{
Number densities of RC stars plotted against distance from the center of the
LMC (left panels) and the SMC (right panels). Top panels show data for the
entire density map, middle panels show data for the northern part and bottom
panels for the southern part of the maps, respectively. The separation between
north and south is with respect to the LMC center (left panels) and the SMC
center (right panels). Light gray crosses represent all 754 subfields listed
in Table~\ref{tab:densities} while black crosses represent subfields in the
conservative MBR area between declinations $-70^{\circ}$ and $-75^{\circ}$.
Dark gray dots show a subset of subfields excluding the SMC and its periphery
(in the case of the LMC, i.e. left panels) or the LMC and its periphery (in the
case of the SMC, i.e. right panels).
Large green circles show median values of gray dots (top panels) while large
purple squares show median values of gray crosses (middle and bottom panels).
The solid line is a radial fit to dark gray dots in top panels
for $r\leq8^{\circ}$ in the form:
$f=f_{0}\times e^{-r/h}$ where $f_{0}$ is the central density and $h$ is the
scale height of the fit. Parameters of the fits are shown on each plot.
The horizontal dotted line marks a median background value (36 stars/deg$^2$)
for RC stars based on 50 subfields in the northern part of the maps, where
we do not expect LMC/SMC stars.
The vertical dashed line at $r=21^{\circ}$ shows the approximate distance
between the LMC and the SMC centers.
}
\label{fig:profiles_rc}
\end{figure*}

The RC number density map shows that intermediate-age stars are fairly
concentrated around the MCs, although there is a distinct asymmetry in the
central and southern part of the map (Figure~\ref{fig:maps_rc}), where
the two populations overlap.
In Figure~\ref{fig:profiles_rc}, we plot number densities of RC stars as a
function of distance from the LMC/SMC center (left/right panels), which in
the case of the LMC is
$\alpha=05^{\rm h}29^{\rm m}$ and $\delta=-69^{\circ}30'$
\citep{vanderMarel2001} and in the case of the SMC is
$\alpha=01^{\rm h}00^{\rm m}31^{\rm s}$ and $\delta=-72^{\circ}43'11''$,
estimated by \cite{Nidever2011} for the outer SMC population.
We immediately see that there is a continuity of RC stars (top panels of
Figure~\ref{fig:profiles_rc}) in the MBR area that significantly exceeds
the median background level indicated by dotted horizontal lines.

To separate stars from the two galaxies, we plot a subset of subfields
excluding the SMC and its periphery (in the case of the LMC, i.e. top left
panel) or the LMC and its periphery (in the case of the SMC, i.e. top right
panel) with dark gray dots, as well as their median values with large green
circles. We see that the RC stars are present as far as $\sim15$~deg from
the LMC center and $\sim9$~deg from the SMC center, although this does not
yet mean that the RC populations of the two galaxies overlap, since both
the LMC and the SMC ellipses have major axes pointing away from the main
MBR. But if we consider subfields only in the conservative Bridge region
between the Clouds
($-75^{\circ} <\delta<-70^{\circ}$), which are plotted with black crosses, we
see no definite distinction as where the LMC population ends and the SMC
population begins, which shows that the RC populations of the two galaxies
overlap, as inferred from the density map (Figure~\ref{fig:maps_rc}).

We also fit radial density profiles to number densities marked with dark gray
dots, in the form: $f=f_{0}\times e^{-r/h}$ where $f_{0}$ is the central
density and $h$ is the scale height of the fit. Resulting parameters
($f_{0},~h)$ for fits in the range $0 \le r \le 8$~deg are marked on the plots.
We do realize that using radial density profile fits for the galaxies with
non-negligible ellipticities (0.1 for the SMC periphery; \citealt{Nidever2011};
and 0.2 for the LMC; \citealt{vanderMarel2001}) results in inaccurate
profile parameters and we would achieve better results with elliptical fits.
This will be addressed in future papers, when the entire SMC and LMC OGLE-IV
data are available. Here we only investigate the qualitative nature
of the population distribution with respect to approximate density
profiles of the Clouds.

There is an evident break in the radial profiles of RC stars at
$r \approx 9$~deg for the LMC and $r \approx 7$~deg for the SMC, already
noticed by \cite{Nidever2011}. These outer populations may be a part of stellar
halos of the two galaxies  and if this is the case, we now see that the two
halos overlap. \cite{Nidever2011} also proposed, that the break population they
observed around the SMC may be a tidal tail or debris, which at small radii
($4-9$~kpc $\approx 2$~tidal radii) would look symmetric, thus making it look
like a classical stellar halo. OGLE-IV data presented in this paper map the
SMC periphery out to $3-4$ tidal radii and no evident tidal tail is visible,
favoring the classical stellar halo explanation. However, we cannot make
more definite conclusions before the entire LMC and SMC data are analyzed.

We further divide the RC density map into northern and southern part (middle
and bottom panels of Figure~\ref{fig:profiles_rc}, respectively) with respect
to the LMC center (left panels) and to the SMC center (right panels), and we
mark median number densities with large purple squares. We clearly see that the
overlap is visible only in the southern part, and practically unobserved in the
northern part of the map, where the number densities follow radial density
profiles of the two galaxies.

\begin{figure*}[htp]
\centering
\vspace{-0.5cm}
\begin{tabular}{cc}
\includegraphics[width=7cm]{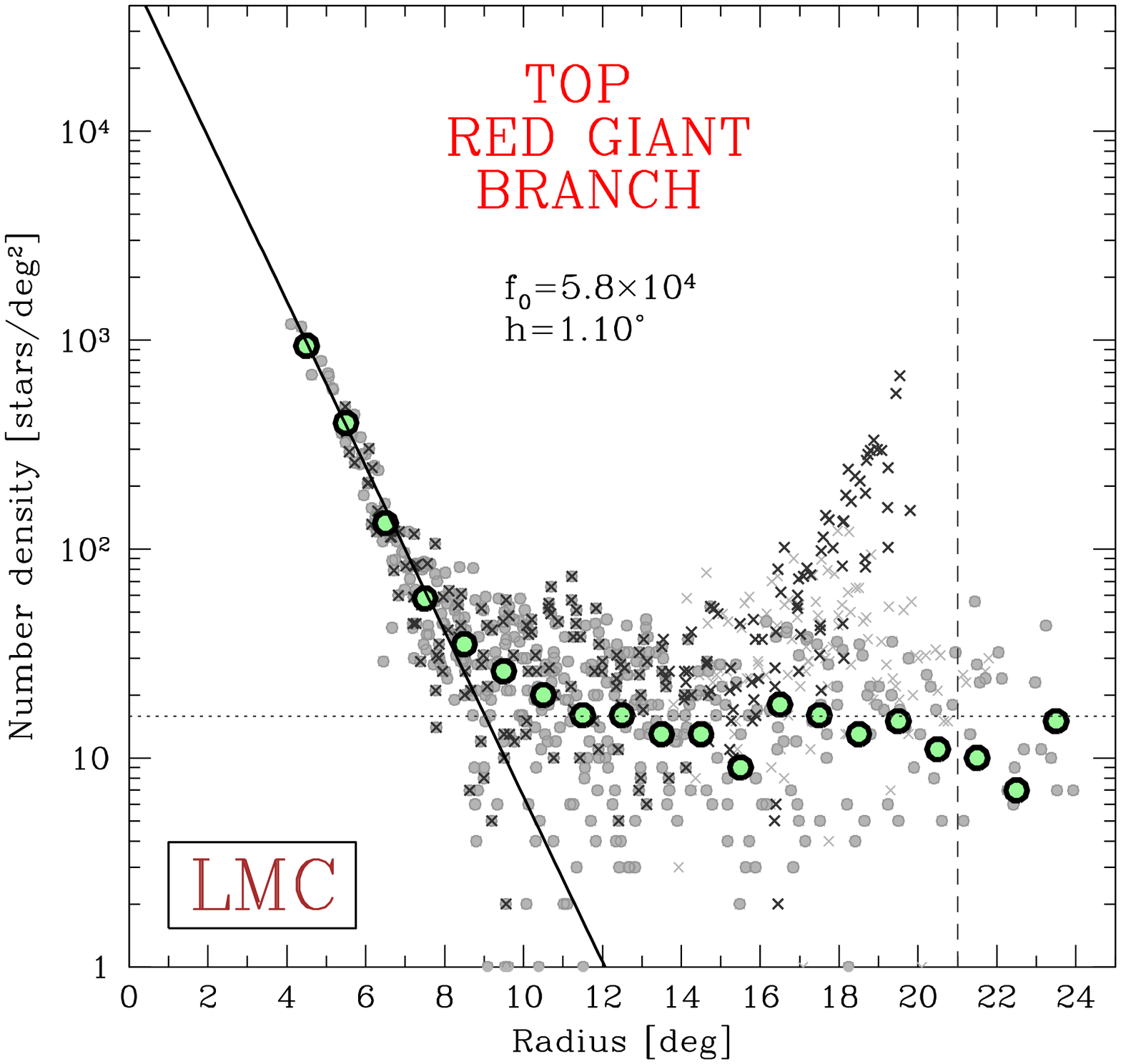} &
\includegraphics[width=7cm]{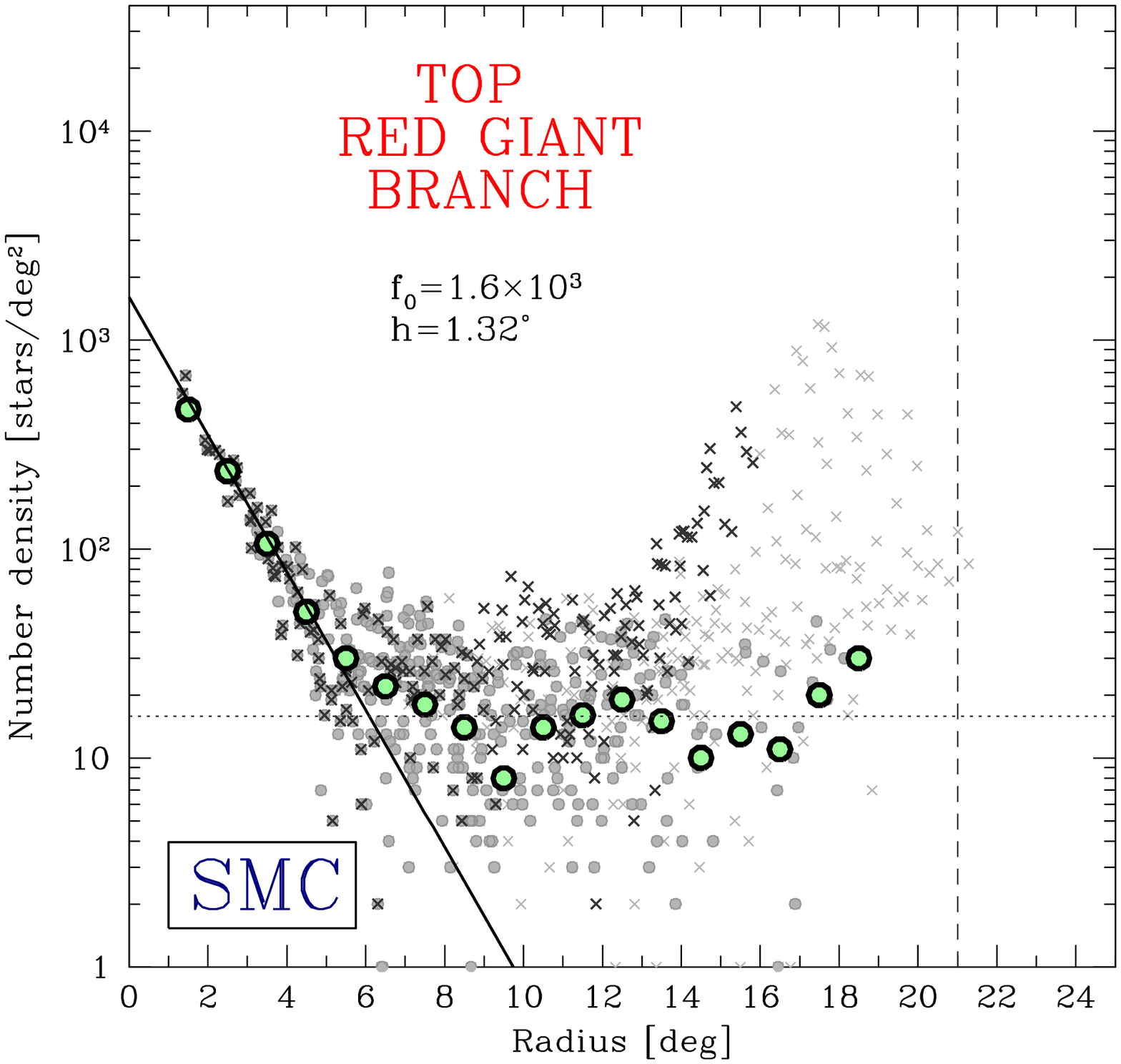} \\ [-4ex]
\includegraphics[width=7cm]{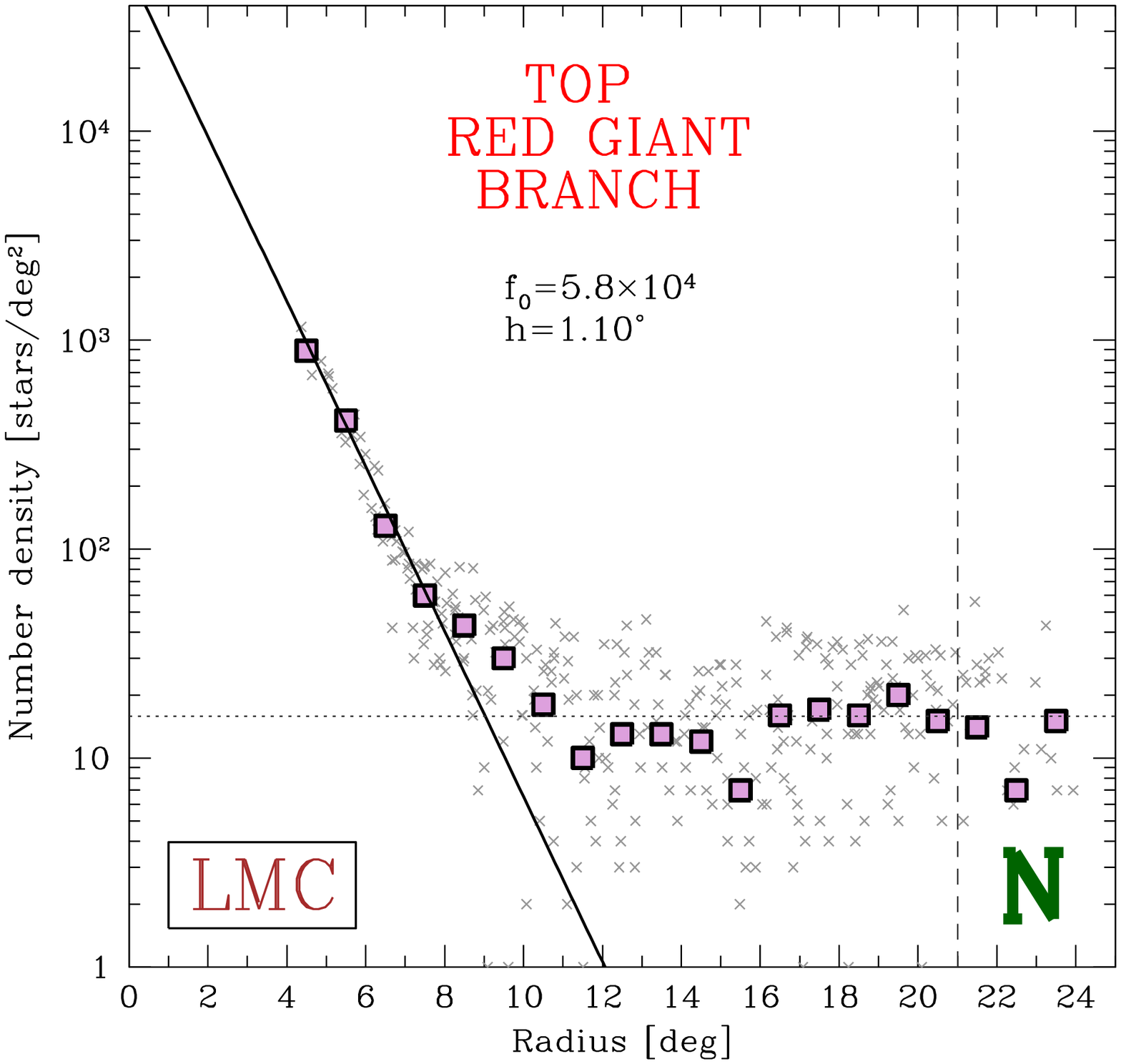} &
\includegraphics[width=7cm]{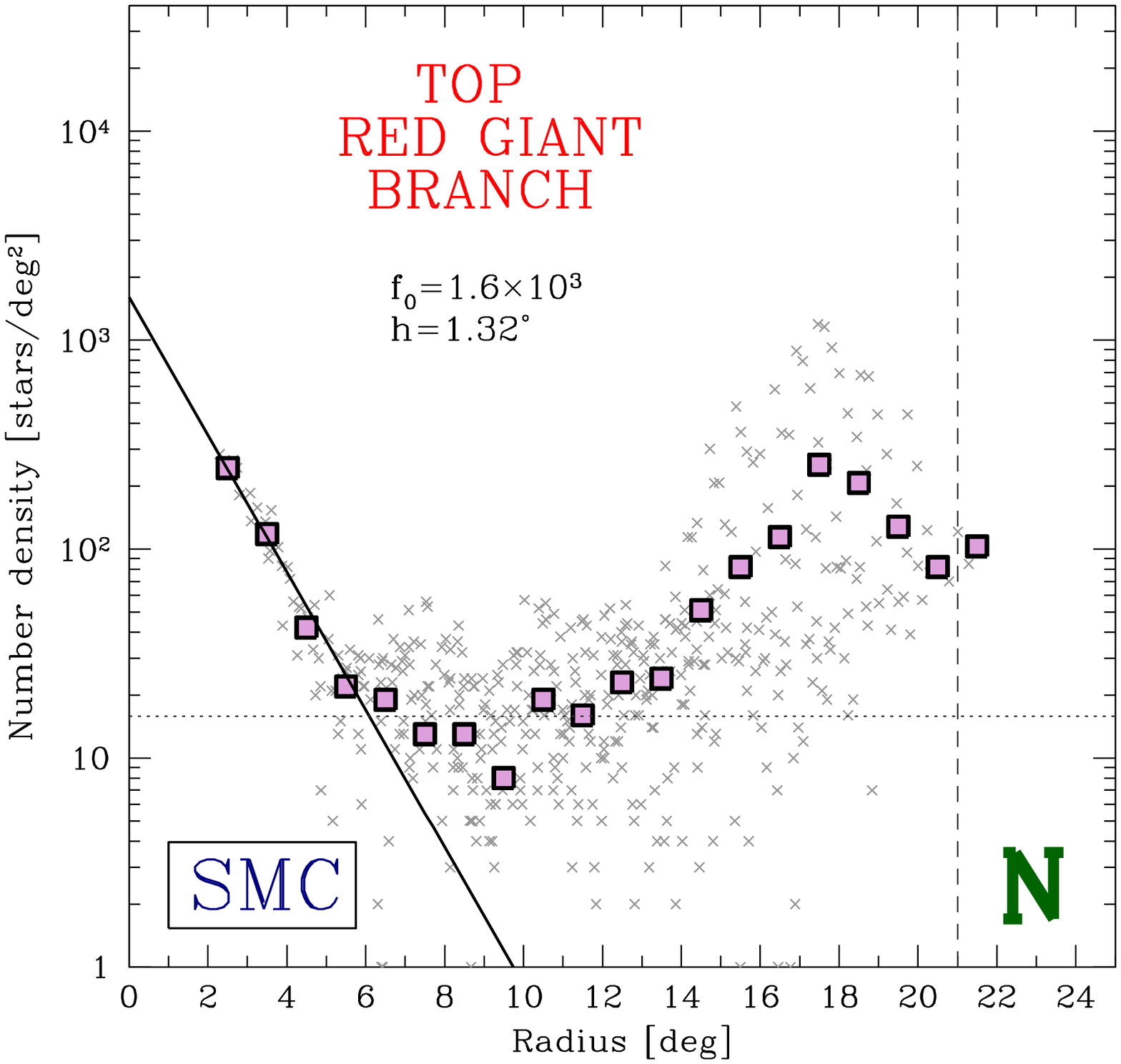} \\ [-4ex]
\includegraphics[width=7cm]{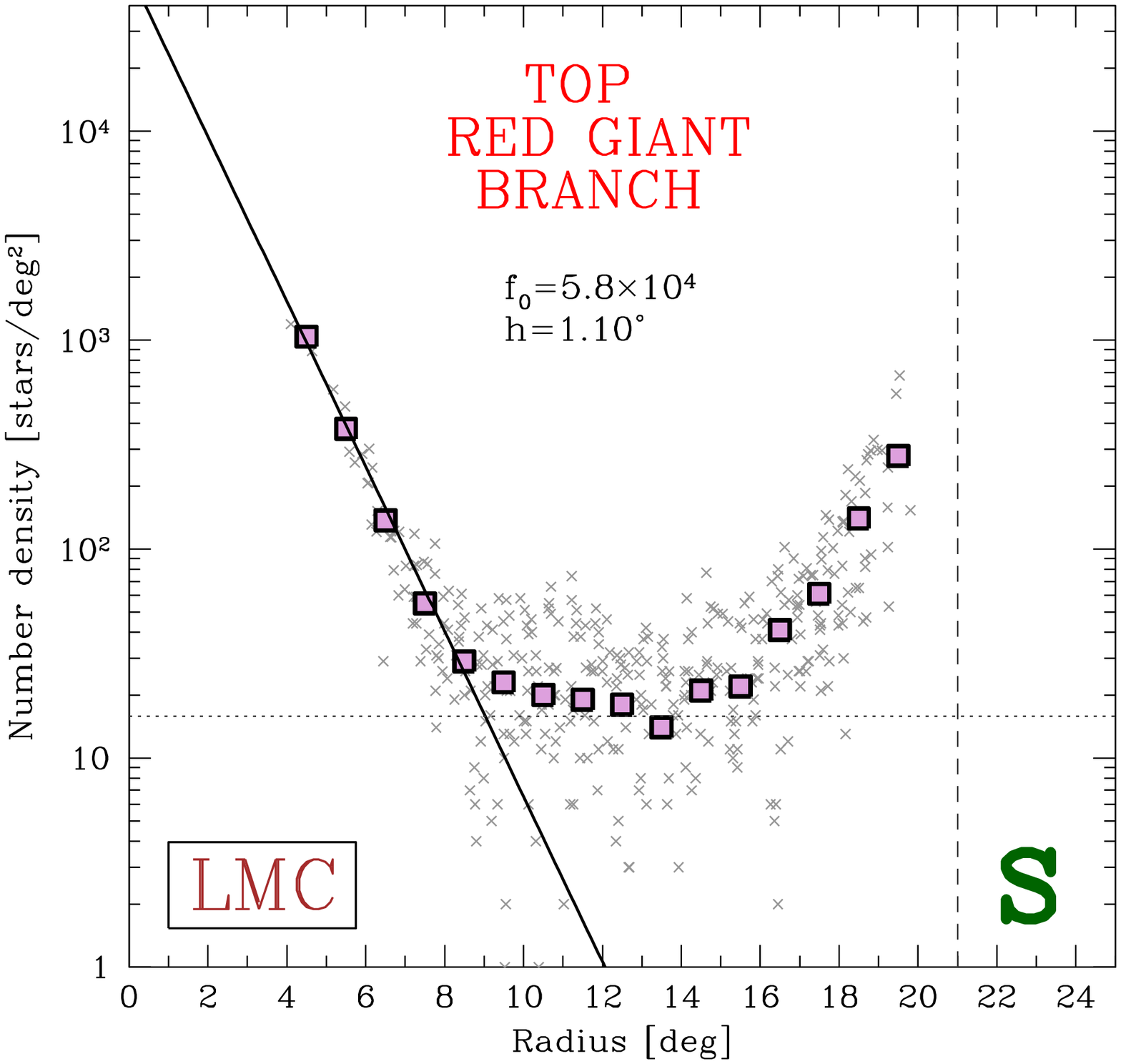} &
\includegraphics[width=7cm]{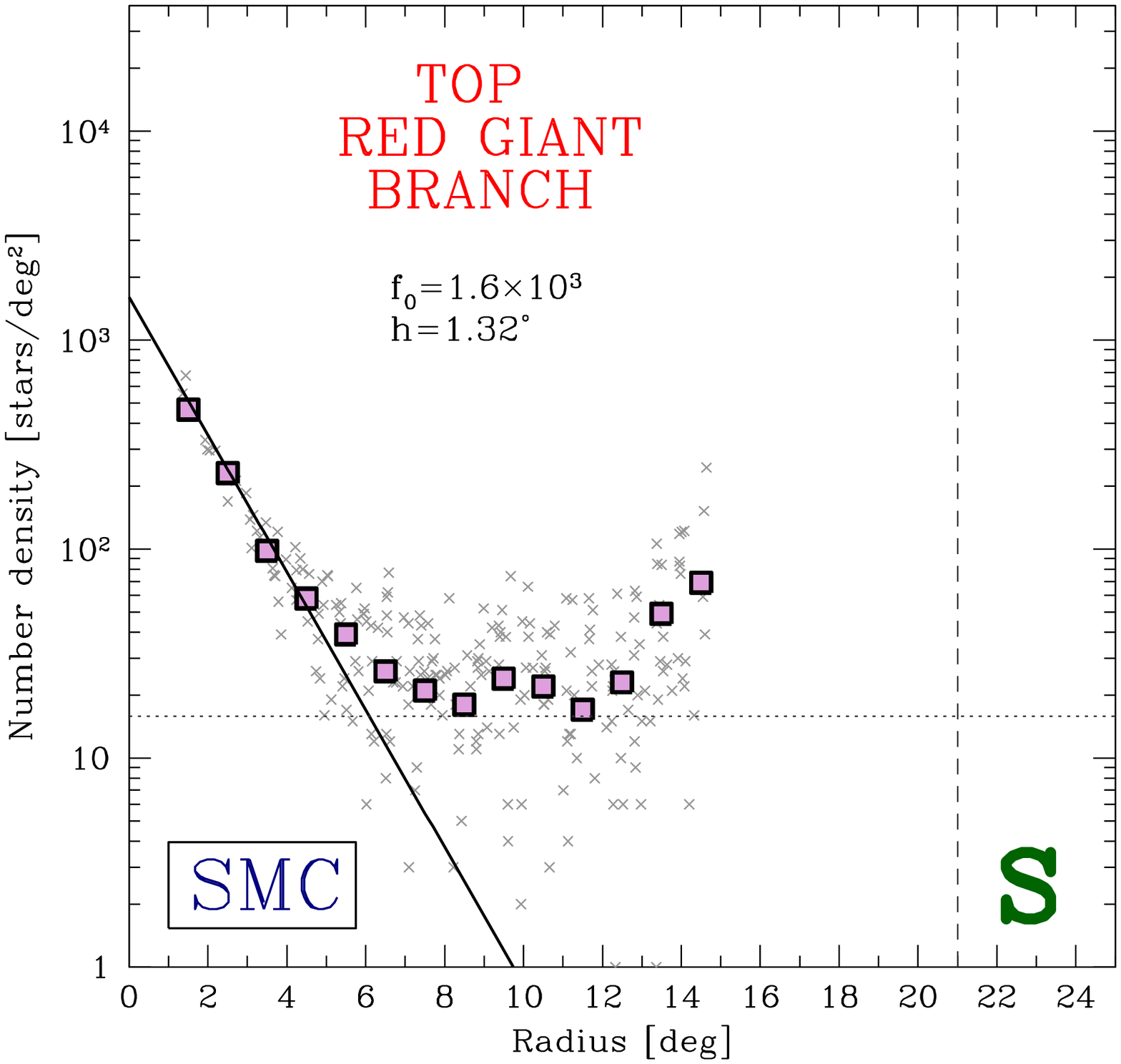} \\
\end{tabular}
\caption{
Number densities of top RGB stars (I<16~mag and V-I<3~mag), plotted against
distance from the center of the LMC (left panels) and the SMC (right panels).
Top panels show data for the
entire density map, middle panels show data for the northern part and bottom
panels for the southern part of the maps, respectively. The separation between
north and south is with respect to the LMC center (left panels) and the SMC
center (right panels). Light gray crosses represent all 754 subfields listed
in Table~\ref{tab:densities} while black crosses represent subfields in the
conservative MBR area between declinations $-70^{\circ}$ and $-75^{\circ}$.
Dark gray dots show a subset of subfields excluding the SMC and its periphery
(in the case of the LMC, i.e. left panels) or the LMC and its periphery (in the
case of the SMC, i.e. right panels).
Large green circles show median values of gray dots (top panels) while large
purple squares show median values of Grey crosses (middle and bottom panels).
The solid line is a radial fit to dark gray dots in top panels for
$r\leq8^{\circ}$ in the form:
$f=f_{0}\times e^{-r/h}$ where $f_{0}$ is the central density and $h$ is the
scale height of the fit. Parameters of the fits are shown on each plot.
The horizontal dotted line marks a median  background value (12 stars/deg$^2$)
for top RGB stars based on 50 subfields in the northern part of the maps, where
we do not expect LMC/SMC stars.
The vertical dashed line at $r=21^{\circ}$ shows the approximate distance
between the LMC and the SMC centers.
}
\label{fig:profiles_rgb_top}
\end{figure*}

Unlike the RC population, the intermediate-age stars represented by young RGB
and AGB stars (Figure~\ref{fig:maps_rgb_top}) do not reach as far into the
Bridge, and there is no asymmetry between the northern and the southern part
of the map, nor a significant diversion from a radial density profile
(Figure~\ref{fig:profiles_rgb_top}), suggesting that these are LMC and SMC stars.
However, we do observe a mildly higher number of stars in the eastern and the
south-western part of the classical Bridge which are fairly consistent with
the results of \cite{Bagheri2013} who identified a population of young RGB and
AGB stars ($\sim400$~Myr to 5~Gyr old) in 2MASS data in that area.
However, our data show a gap between the galaxies rather than a continuous
stream of stars seen by 2MASS, which may further support the hypothesis that
these are genuine LMC/SMC stars rather than a tidal stream.

Finally, in Figures~\ref{fig:maps_rgb_bot} and~\ref{fig:profiles_rgb_bot} we
show a number density map and radial density profiles, respectively, of an old
population represented by the bottom part of the RGB. All designations are the
same as in previous figures. Old stars do not extend as far into the MBR as the
intermediate-age (RC) stars, and are observed out to $\sim11$~deg from the LMC
center and $\sim7$~deg from the SMC center, which most probably means that
these are stellar halos of the two galaxies. Also their distribution is more
symmetrical, although there is an overdensity in the southern part of the map,
also visible as a deviation from a radial density profile (bottom panels in
Figure~\ref{fig:profiles_rgb_bot}).

\section{Summary}

In this paper, we analyzed OGLE-IV observations of the entire Magellanic Bridge
region fully covering over 270~deg$^2$ between the Magellanic Clouds (1.3~h
$\leq \alpha \leq 4.65$~h, $-80$~deg $\leq \delta \leq -63.5$~deg).
This unique dataset allowed us to construct detailed number density maps for
three key stellar populations: the young stars, and the intermediate-age and
old populations, represented by the red clump and the red giant branch stars.

\begin{figure*}[htp]
\centerline{\includegraphics[width=18.5cm]{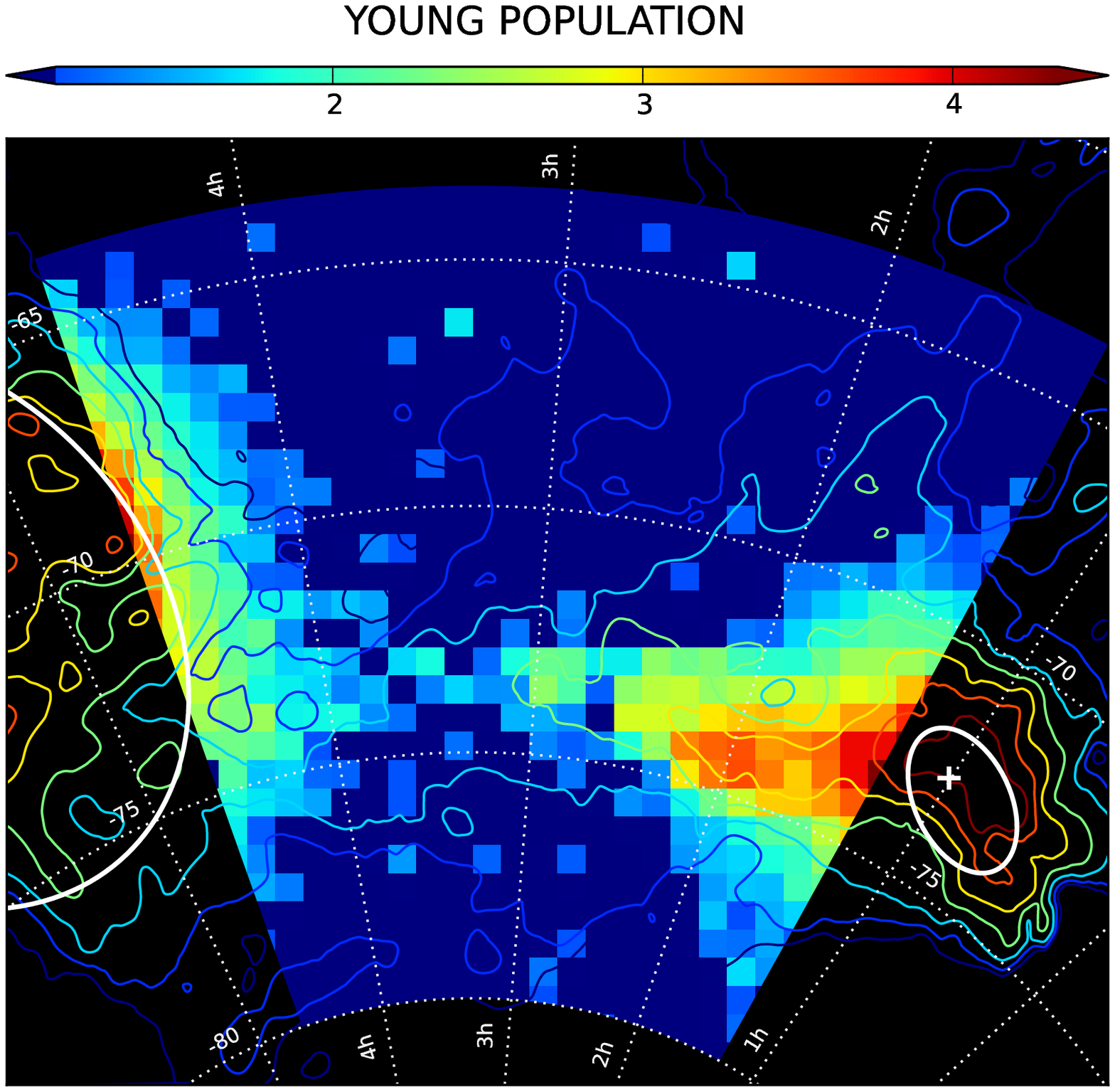}}
\caption{
A spatial density map of the Young Population stars in the Magellanic Bridge
region, in a Hammer equal-area projection centered at $\alpha=3.3$~h and
$\delta=-70$~deg. The color-coded value of each ``pixel'' is a logarithm of
the number of stars per square degree area (as indicated on the color bar at
the top of the figure), while each ``pixel'' area is $\sim 0.335\deg^2$.
Note the fairly isolated ``OGLE island'' mid-way between the Clouds at
$\alpha \approx 2.9$~h and $\delta \approx-73.5$~deg.
A median background level was estimated from 40 most southern fields to be
1~star/deg$^2$ with a standard deviation of 6~stars/deg$^2$. Detections
weaker than $2\sigma$ above the median background level have been given the
background color. All number densities are listed in Table~\ref{tab:densities}.
An approximate location of the LMC disk and the main stellar body of the SMC
are marked with white ellipses centered at $\alpha=05^{\rm h}29^{\rm m}$, 
$\delta=-69^{\circ}30'$, and $\alpha=00^{\rm h}54^{\rm m}$,
$\delta=-72^{\circ}57'$, respectively.
The white cross marks the SMC center of the outer SMC population found by
\cite{Nidever2011} at $\alpha=01^{\rm h}00^{\rm m}31^{\rm s}$ and
$\delta=-72^{\circ}43'11''$.
Color contours mark neutral hydrogen (\HI) emission integrated over the velocity
range $80 < v < 400$~km/s, where each contour represents the \HI~column density
twice as large as the neighboring contour. \HI~column densities are in the
range $10^{20}-4\times10^{21} {\rm cm}^{-2}$.
Data were taken from the LAB survey of Galactic \HI~\citep{Kalberla2005}.
}
\label{fig:maps_yp}
\end{figure*}

The density map for the youngest population (Figure~\ref{fig:maps_yp}) confirms
that the majority of young stars are found in the western part of the classical
Bridge, but what is more important, shows that the young population is also
present in the eastern part of the classical Bridge region, which was not
observed before. Even though number densities in that region are much lower
than in the western Bridge, they are still 4 to $14\sigma$ detections.
This means that there is a continuous stream of young stars connecting the
two galaxies.

The density map also shows detailed distribution of the YP that reveals two
overdensities in the western Bridge, one close to the SMC (R.A.~$\approx 2$~h,
Dec~$\approx-74$~deg), and the other, fairly isolated and located
approximately mid-way between the Clouds (R.A.~$\approx 2.9$~h,
Dec~$\approx-73.5$~deg), which we call the OGLE island. We show that these
overdensities are well matched by \HI~surface density contours, including
the OGLE island. On the other hand, the YP in the eastern Bridge is slightly
offset ($\sim 2$~deg north) from the highest density \HI~ridge, and is
concentrated around $\delta \approx -73.5$~deg -- a continuation of the OGLE
island toward the LMC.

The density map for the top RGB and AGB stars that represent an
intermediate-age population younger that the RC but older than the YP, show a
rather symmetric distribution around the Magellanic Clouds which suggests that
this may simply be the extent of the galaxies into the Magellanic Bridge. On
the other hand, the density map for the intermediate-age population represented
by RC stars shows that there is a continuity of RC stars between the Clouds
in the central and southern part of the Magellanic Bridge, and it is reflected
by a deviation from the radial density profile. This may indicate that this is
indeed a tidal stellar component of the MBR.

\begin{figure*}[htp]
\centerline{\includegraphics[width=18.5cm]{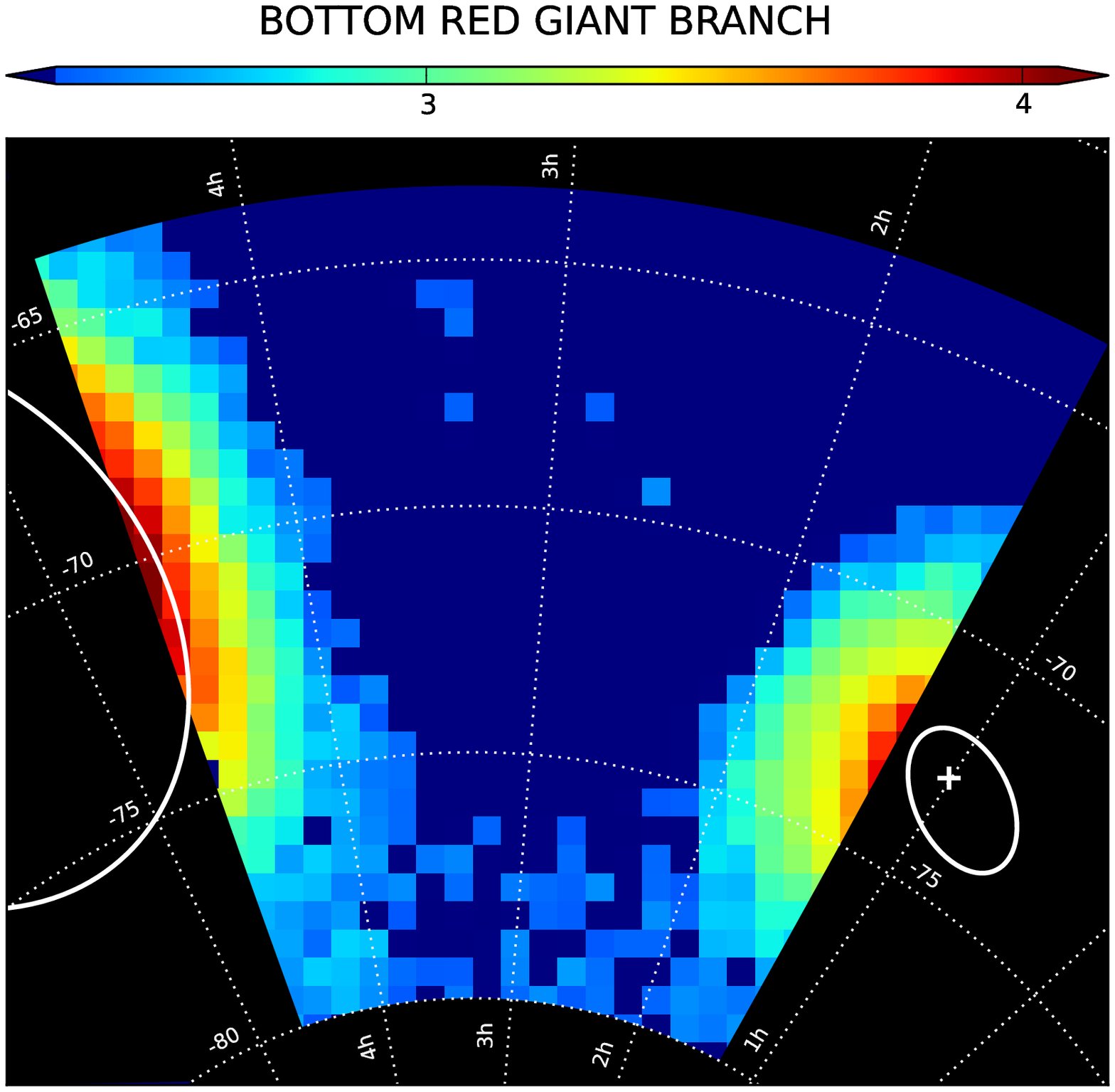}}
\caption{
Spatial density maps of the bottom part of the Red Giant Branch stars
(I>19~mag and I<20~mag) in the Magellanic Bridge
region, in a Hammer equal-area projection centered at $\alpha=3.3$~h and
$\delta=-70$~deg. The color-coded value of each ``pixel'' is a logarithm of
the number of stars per square degree area (as indicated on the color bar at
the top of the figure), while each ``pixel'' area is $\sim 0.335\deg^2$.
A median background level was estimated from 50 northern fields to be
112~stars/deg$^2$ with a standard deviation of 65~stars/deg$^2$. Detections
weaker than $2\sigma$ above the median background level have been given the
background color. All number densities are listed in Table~\ref{tab:densities}.
An approximate location of the LMC disk and the main stellar body of the SMC
are marked with white ellipses centered at $\alpha=05^{\rm h}29^{\rm m}$,
$\delta=-69^{\circ}30'$, and $\alpha=00^{\rm h}54^{\rm m}$,
$\delta=-72^{\circ}57'$, respectively.
The white cross marks the SMC center of the outer SMC population found by
\cite{Nidever2011} at $\alpha=01^{\rm h}00^{\rm m}31^{\rm s}$ and
$\delta=-72^{\circ}43'11''$.
}
\label{fig:maps_rgb_bot}
\end{figure*}

\begin{figure*}[htp]
\centering
\vspace{-0.5cm}
\begin{tabular}{cc}
\includegraphics[width=7cm]{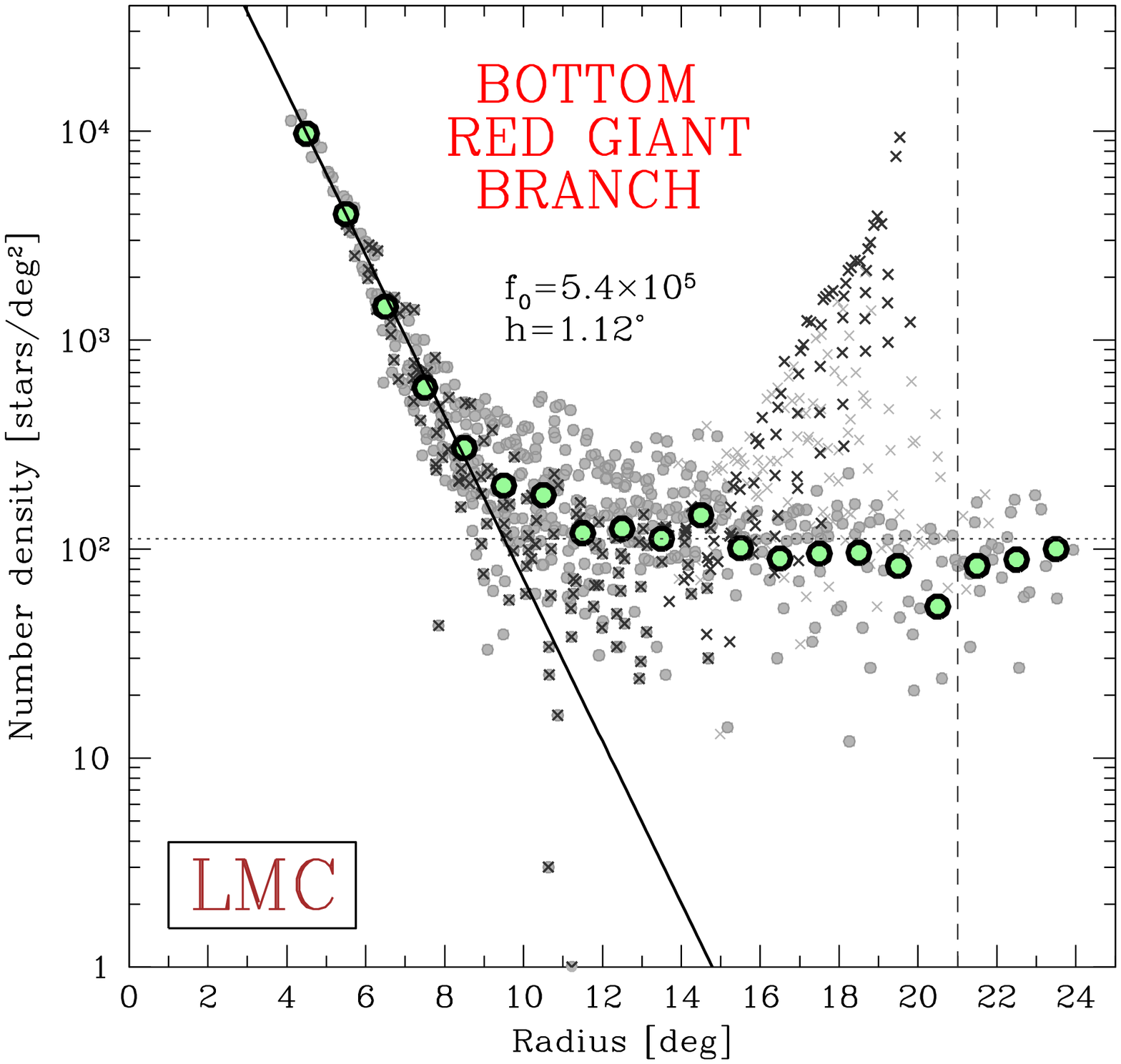} &
\includegraphics[width=7cm]{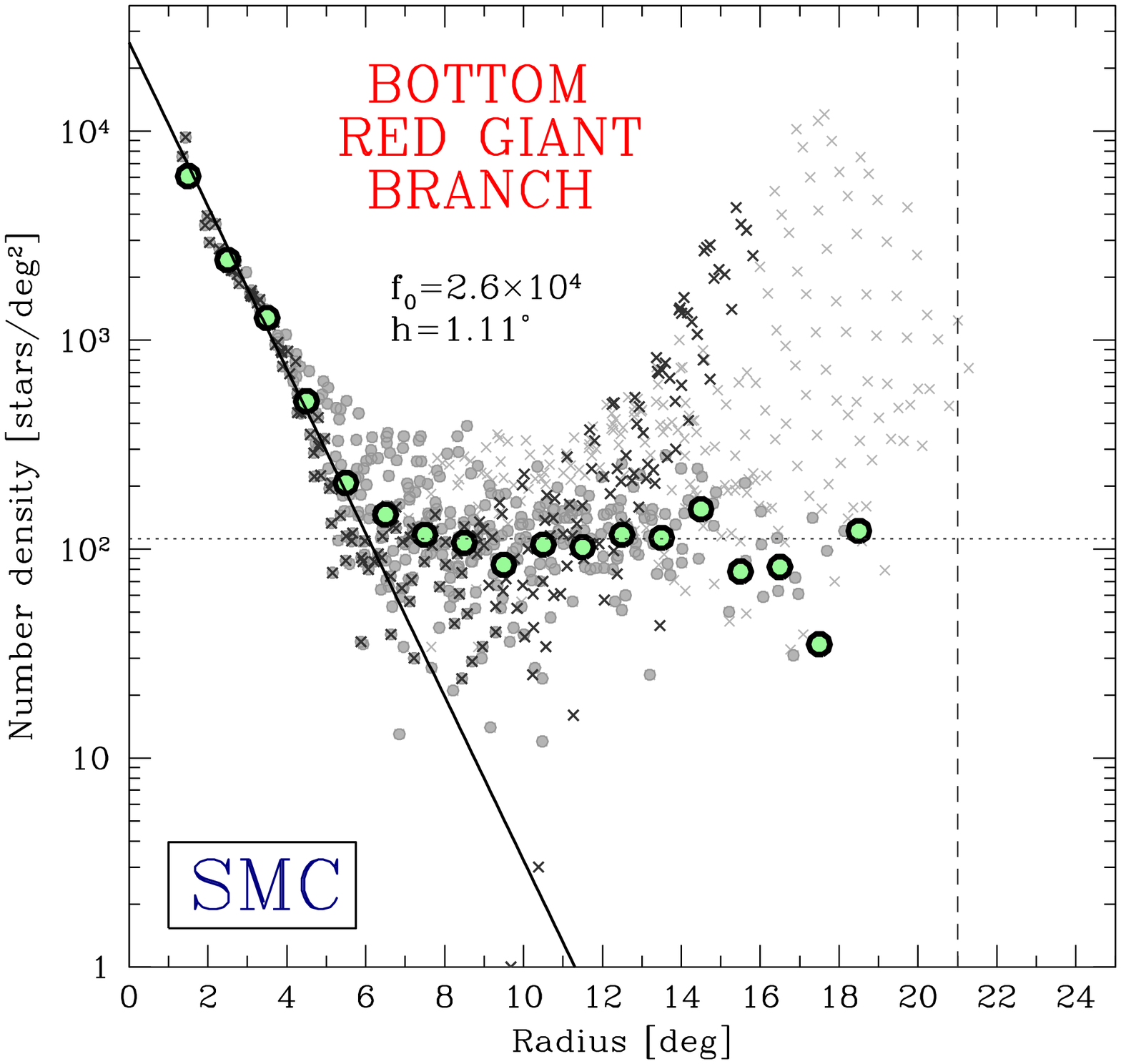} \\ [-4ex]
\includegraphics[width=7cm]{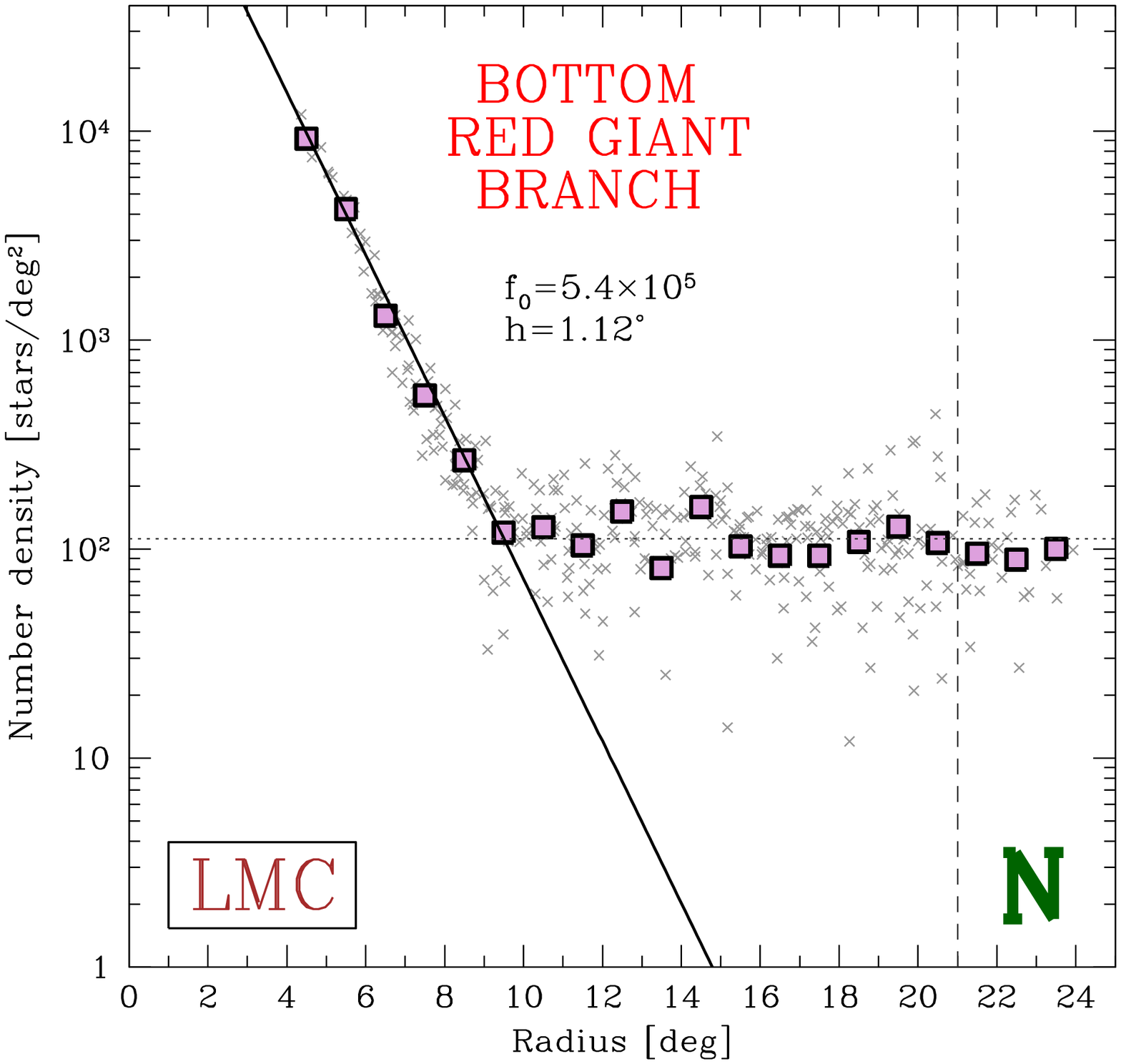} &
\includegraphics[width=7cm]{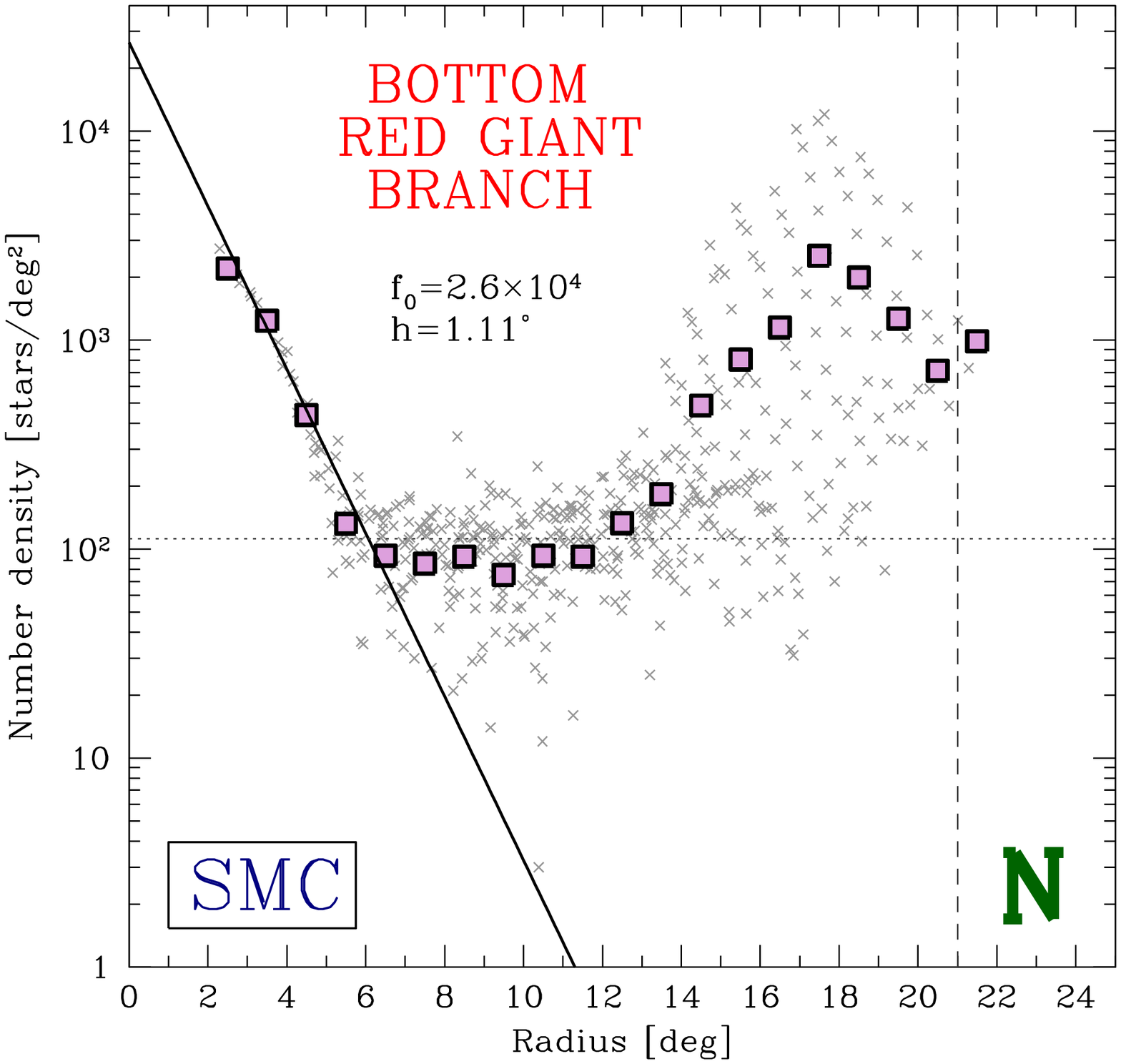} \\ [-4ex]
\includegraphics[width=7cm]{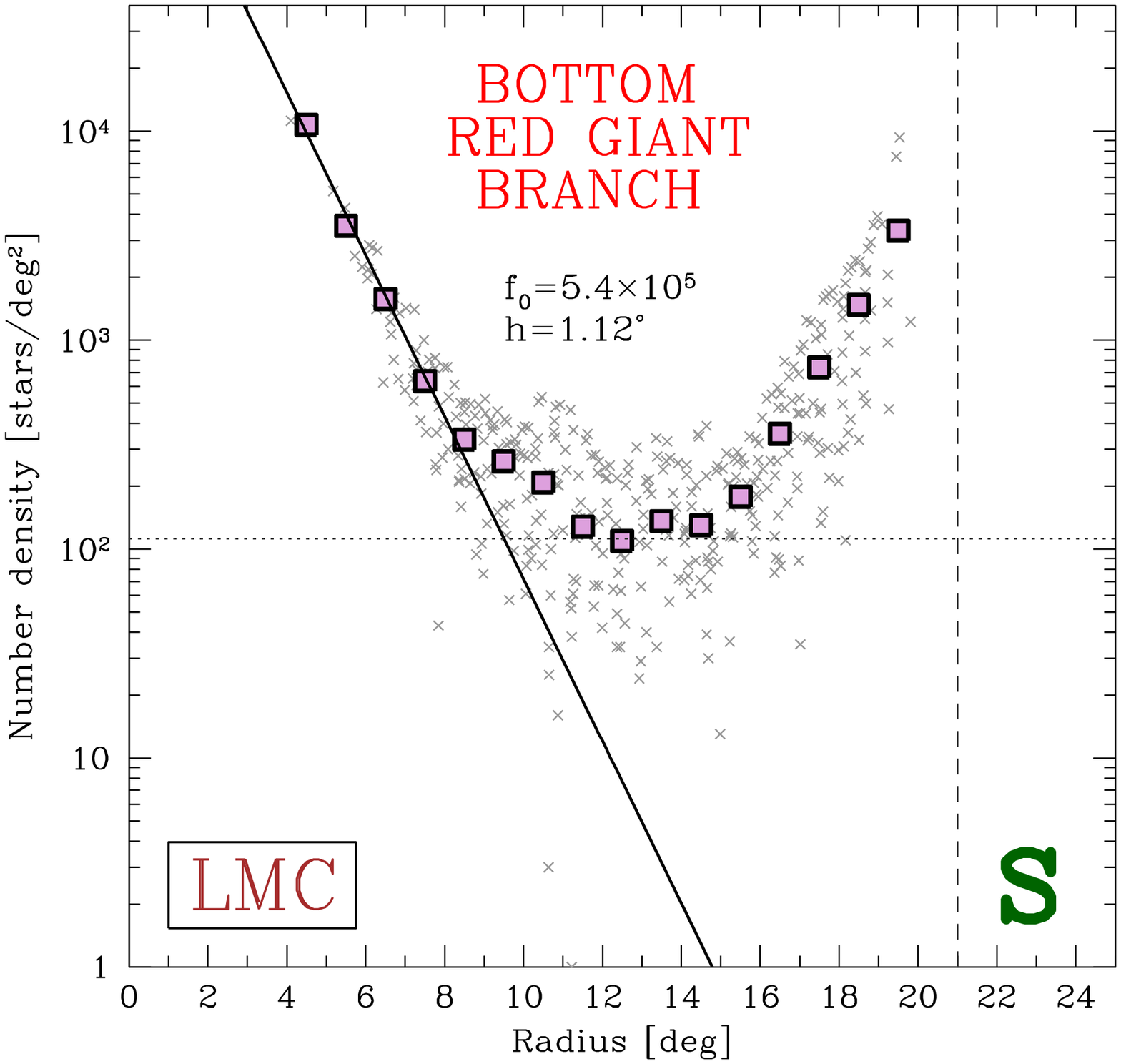} &
\includegraphics[width=7cm]{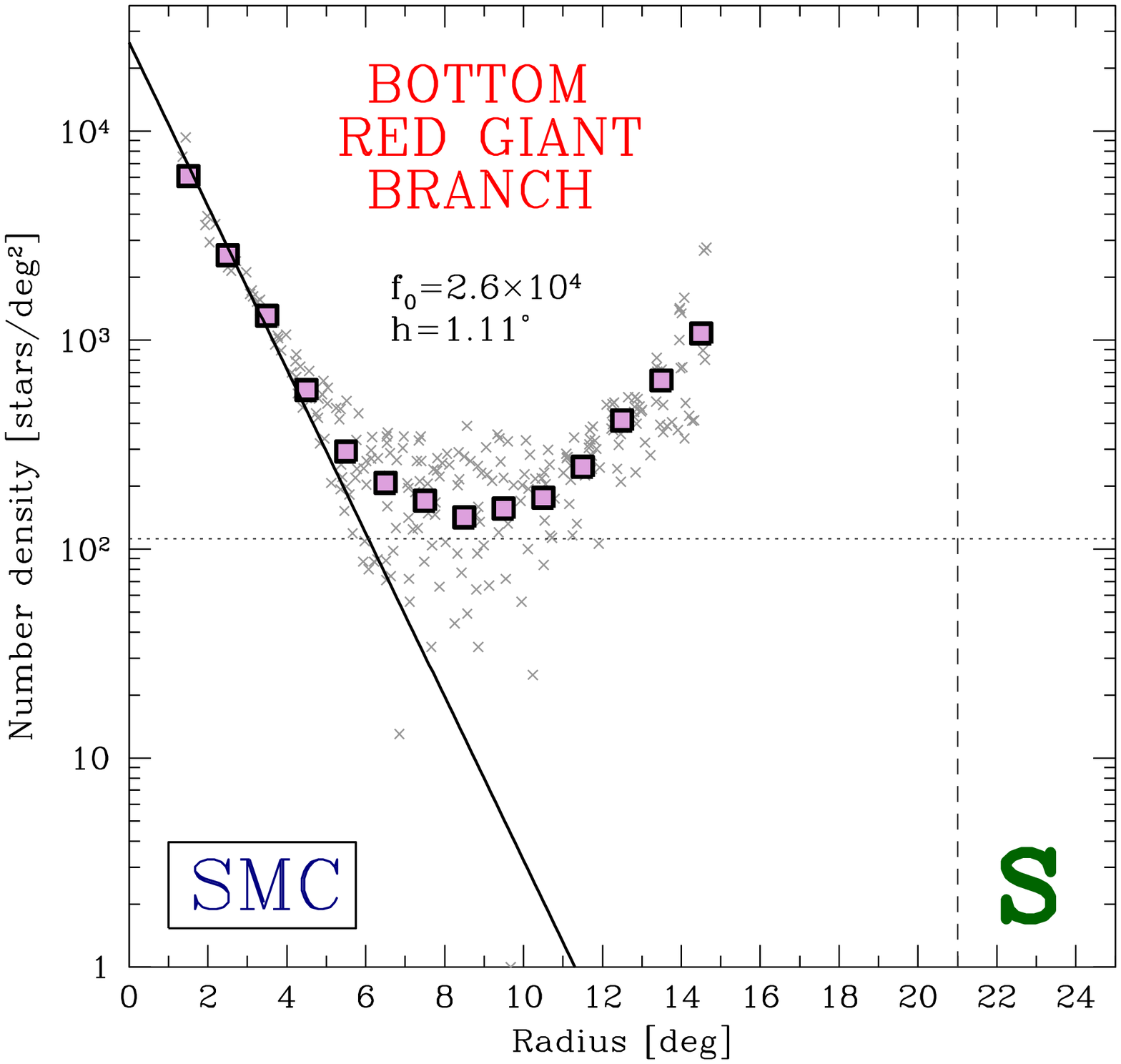} \\
\end{tabular}
\caption{
Number densities of bottom RGB stars (I<19~mag and I>20~mag), plotted against 
distance from the center of the LMC (left panels) and the SMC (right panels). 
Top panels show data for the
entire density map, middle panels show data for the northern part and bottom
panels for the southern part of the maps, respectively. The separation between
north and south is with respect to the LMC center (left panels) and the SMC
center (right panels). Light gray crosses represent all 754 subfields listed
in Table~\ref{tab:densities} while black crosses represent subfields in the
conservative MBR area between declinations $-70^{\circ}$ and $-75^{\circ}$.
Dark gray dots show a subset of subfields excluding the SMC and its periphery
(in the case of the LMC, i.e. left panels) or the LMC and its periphery (in the
case of the SMC, i.e. right panels).
Large green circles show median values of gray dots (top panels) while large
purple squares show median values of gray crosses (middle and bottom panels).
The solid line is a radial fit to dark gray dots in top panels for
$r\leq8^{\circ}$ in the form:
$f=f_{0}\times e^{-r/h}$ where $f_{0}$ is the central density and $h$ is the
scale height of the fit. Parameters of the fits are shown on each plot.
The horizontal dotted line marks a median background value (112 stars/deg$^2$)
for bottom RGB stars based on 50 subfields in the northern part of the maps, where
we do not expect LMC/SMC stars.
The vertical dashed line at $r=21^{\circ}$ shows the approximate distance
between the LMC and the SMC centers.
}
\label{fig:profiles_rgb_bot}
\end{figure*}

We observe only minor mixing of the old populations of the LMC and SMC,
represented by the bottom part of the RGB, in the southern part of the MBR, and
this is visible both on the number density map in Figure~\ref{fig:maps_rgb_bot}
as well as in number density profiles in Figure~\ref{fig:profiles_rgb_bot},
where only a slight break in the radial density profile is present in the
southern part of the MBR. The fact that this population is present only in the
southern Bridge suggests that what we see are the halos of the two galaxies.

The distribution of the young population is consistent with \HI~densities, as it
was formed within the MBR after gas was drawn out of the SMC in a past
interaction between the Clouds $\sim250$~Myr ago. It is expected that this
interaction would also tidally strip stars, and such low metallicity candidates
have been observed in the outskirts of the LMC (e.g. \citealt{Olsen2011}).
Our density maps clearly show presence of intermediate-age stars in the MBR
that could not have been formed there. However, the fact that their distribution
is far from uniform and appears as a break in radial density profiles of the
galaxies, is a bit puzzling. In addition, we observe this population only in the
southern, and to some extent in the central, parts of the Bridge, meaning that
it is strongly offset from the gaseous component of the MBR.
If it was a dislocated tidal stream of stars between the two galaxies it would
mean that there was an additional process that shaped the current Magellanic
System, such as ram pressure stripping from our Galaxy (\citealt{Diaz2012},
\citealt{Noel2013}). On the other hand, this should affect old stars as well as
intermediate-age stars, and in the case of old stars there is no evident
overlap of the LMC and SMC populations, although we do see some detections
of an old population in the south of the maps. So it is possible, that what we
observe on RC density maps are rather overlapping halos of the MCs, than a
tidally stripped component. However, this will have to be verified by further
analysis of the entire OGLE-IV data of the Magellanic Clouds and their
surroundings.

Presented number density maps form a first uniform dataset on stellar
populations in the area between the Magellanic Clouds, much larger than the
classical Magellanic Bridge. This is a unique database that may be used for
testing models and simulations of past interaction between the Magellanic
Clouds and the Milky Way. Data used to make these density maps are available
in an electronic form from the OGLE Internet archive
http://ogle.astrouw.edu.pl.

\acknowledgements

D.M.S. is supported by the Polish National Science Center (NCN) under the grant
no. 2013/11/D/ST9/03445. A.M.J. is supported by the Polish Ministry of Science
and Higher Education Diamond Grant no. 0148/DIA/2014/43. The OGLE project has
received funding from the European Research Council under the European
Community's Seventh Framework Programme (FP7/2007-2013) / ERC grant agreement
no. 246678 to AU.

This work has made use of NASA's Astrophysics Data System Bibliographic
Services.


\setlength{\tabcolsep}{2pt}
\setlength{\tabcolsep}{6pt}
\notetoeditor{(fig:profiles_rc) black and white for print, in color only in
the electronic version}

\setlength{\tabcolsep}{2pt}
\setlength{\tabcolsep}{6pt}
\notetoeditor{(fig:profiles_rgb_top) black and white for print, in color only in
the electronic version}

\setlength{\tabcolsep}{2pt}
\setlength{\tabcolsep}{6pt}
\notetoeditor{(fig:profiles_rgb_bot) black and white for print, in color only in
the electronic version}

\end{document}

%% file: table1e.tex

\begin{deluxetable*}{ccccc|ccccc|ccccc}
\tablecaption{\footnotesize OGLE-IV field center coordinates in the Magellanic Bridge region in equatorial and galactic coordinates.
\label{tab:fields}}\\
\tablehead{ 
Field  & RA [h] & Dec [deg] & l [deg] & b [deg] &  Field  & RA [h] & Dec [deg] & l [deg] & b [deg] &  Field  & RA [h] & Dec [deg] & l [deg] & b [deg]
}
\startdata
MBR100  & 1.860 & -70.06 & 295.58 & -46.20 &  	MBR144  & 4.031 & -75.00 & 289.16 & -36.57 &  	MBR188  & 2.333 & -63.90 & 287.81 & -50.55 \\
MBR101  & 1.851 & -71.29 & 296.22 & -45.06 &  	MBR145  & 4.095 & -71.92 & 285.75 & -38.04 &  	MBR189  & 2.583 & -69.44 & 290.39 & -45.08 \\
MBR102  & 1.835 & -72.52 & 296.87 & -43.91 &  	MBR146  & 4.139 & -73.15 & 286.93 & -37.22 &  	MBR190  & 2.583 & -68.21 & 289.41 & -46.11 \\
MBR103  & 1.816 & -73.75 & 297.50 & -42.76 &  	MBR147  & 1.847 & -68.83 & 295.08 & -47.38 &  	MBR191  & 2.583 & -66.98 & 288.40 & -47.12 \\
MBR104  & 1.794 & -74.99 & 298.11 & -41.61 &  	MBR148  & 1.803 & -67.60 & 294.81 & -48.62 &  	MBR192  & 2.583 & -65.75 & 287.35 & -48.13 \\
MBR105  & 1.768 & -76.22 & 298.69 & -40.46 &  	MBR149  & 1.761 & -66.37 & 294.55 & -49.87 &  	MBR193  & 2.583 & -64.52 & 286.25 & -49.12 \\
MBR106  & 2.100 & -70.67 & 294.30 & -45.19 &  	MBR150  & 1.722 & -65.14 & 294.29 & -51.11 &  	MBR194  & 2.583 & -63.29 & 285.11 & -50.11 \\
MBR107  & 2.100 & -71.90 & 294.99 & -44.06 &  	MBR151  & 1.687 & -63.91 & 294.00 & -52.35 &  	MBR195  & 2.833 & -70.06 & 289.41 & -43.83 \\
MBR108  & 2.100 & -73.14 & 295.66 & -42.92 &  	MBR152  & 2.119 & -69.45 & 293.45 & -46.26 &  	MBR196  & 2.833 & -68.83 & 288.35 & -44.80 \\
MBR109  & 2.100 & -74.37 & 296.30 & -41.79 &  	MBR153  & 2.061 & -68.22 & 293.12 & -47.51 &  	MBR197  & 2.833 & -67.59 & 287.26 & -45.76 \\
MBR110  & 2.100 & -75.60 & 296.92 & -40.65 &  	MBR154  & 2.007 & -66.98 & 292.80 & -48.76 &  	MBR198  & 2.833 & -66.36 & 286.12 & -46.71 \\
MBR111  & 2.349 & -71.29 & 293.14 & -44.10 &  	MBR155  & 1.955 & -65.75 & 292.49 & -50.02 &  	MBR199  & 2.833 & -65.13 & 284.95 & -47.64 \\
MBR112  & 2.365 & -72.52 & 293.85 & -42.98 &  	MBR156  & 1.908 & -64.52 & 292.17 & -51.27 &  	MBR200  & 2.833 & -63.90 & 283.74 & -48.57 \\
MBR113  & 2.384 & -73.75 & 294.53 & -41.85 &  	MBR157  & 1.865 & -63.29 & 291.83 & -52.51 &  	MBR201  & 3.083 & -69.44 & 287.49 & -43.46 \\
MBR114  & 2.406 & -74.99 & 295.18 & -40.71 &  	MBR158  & 2.046 & -79.91 & 299.10 & -36.69 &  	MBR202  & 3.083 & -68.21 & 286.33 & -44.37 \\
MBR115  & 2.432 & -76.22 & 295.82 & -39.58 &  	MBR159  & 1.955 & -78.67 & 298.89 & -37.93 &  	MBR203  & 3.083 & -66.98 & 285.13 & -45.26 \\
MBR116  & 2.596 & -70.67 & 291.26 & -44.01 &  	MBR160  & 1.878 & -77.44 & 298.68 & -39.17 &  	MBR204  & 3.083 & -65.75 & 283.89 & -46.13 \\
MBR117  & 2.606 & -71.91 & 292.13 & -42.95 &  	MBR161  & 2.461 & -79.29 & 297.50 & -36.80 &  	MBR205  & 3.083 & -64.52 & 282.62 & -47.00 \\
MBR118  & 2.640 & -73.14 & 292.85 & -41.83 &  	MBR162  & 2.331 & -78.06 & 297.26 & -38.08 &  	MBR206  & 3.083 & -63.29 & 281.30 & -47.85 \\
MBR119  & 2.679 & -74.37 & 293.54 & -40.70 &  	MBR163  & 2.221 & -76.83 & 297.02 & -39.35 &  	MBR207  & 3.333 & -70.06 & 286.81 & -42.12 \\
MBR120  & 2.725 & -75.60 & 294.22 & -39.57 &  	MBR164  & 2.708 & -78.06 & 295.95 & -37.52 &  	MBR208  & 3.333 & -68.83 & 285.59 & -42.96 \\
MBR121  & 2.845 & -71.29 & 290.38 & -42.81 &  	MBR165  & 2.667 & -76.83 & 295.28 & -38.64 &  	MBR209  & 3.333 & -67.59 & 284.34 & -43.79 \\
MBR122  & 2.871 & -72.52 & 291.25 & -41.75 &  	MBR166  & 2.933 & -79.91 & 296.53 & -35.62 &  	MBR210  & 3.333 & -66.36 & 283.06 & -44.60 \\
MBR123  & 2.924 & -73.76 & 291.99 & -40.63 &  	MBR167  & 2.967 & -78.67 & 295.56 & -36.58 &  	MBR211  & 3.333 & -65.13 & 281.73 & -45.40 \\
MBR124  & 2.985 & -74.99 & 292.71 & -39.50 &  	MBR168  & 3.083 & -77.44 & 294.28 & -37.35 &  	MBR212  & 3.333 & -63.90 & 280.37 & -46.19 \\
MBR125  & 3.057 & -76.22 & 293.41 & -38.36 &  	MBR169  & 3.392 & -79.29 & 294.87 & -35.33 &  	MBR213  & 3.583 & -69.44 & 285.04 & -41.55 \\
MBR126  & 3.093 & -70.68 & 288.57 & -42.51 &  	MBR170  & 3.392 & -78.06 & 293.84 & -36.24 &  	MBR214  & 3.583 & -68.21 & 283.75 & -42.31 \\
MBR127  & 3.102 & -71.91 & 289.63 & -41.56 &  	MBR171  & 3.392 & -76.83 & 292.79 & -37.14 &  	MBR215  & 3.583 & -66.98 & 282.42 & -43.06 \\
MBR128  & 3.146 & -73.14 & 290.51 & -40.51 &  	MBR172  & 3.700 & -79.91 & 294.67 & -34.31 &  	MBR216  & 3.583 & -65.75 & 281.06 & -43.80 \\
MBR129  & 3.219 & -74.38 & 291.28 & -39.38 &  	MBR173  & 3.700 & -78.67 & 293.57 & -35.15 &  	MBR217  & 3.583 & -64.52 & 279.67 & -44.51 \\
MBR130  & 3.304 & -75.61 & 292.03 & -38.25 &  	MBR174  & 3.700 & -77.44 & 292.45 & -35.97 &  	MBR218  & 3.583 & -63.29 & 278.24 & -45.21 \\
MBR131  & 3.342 & -71.30 & 287.97 & -41.23 &  	MBR175  & 4.116 & -79.29 & 293.23 & -33.84 &  	MBR219  & 3.833 & -70.06 & 284.66 & -40.14 \\
MBR132  & 3.368 & -72.53 & 289.02 & -40.28 &  	MBR176  & 4.076 & -78.06 & 292.11 & -34.66 &  	MBR220  & 3.833 & -68.83 & 283.33 & -40.84 \\
MBR133  & 3.430 & -73.76 & 289.92 & -39.22 &  	MBR177  & 4.043 & -76.83 & 290.97 & -35.47 &  	MBR221  & 3.833 & -67.59 & 281.97 & -41.52 \\
MBR134  & 3.525 & -74.99 & 290.72 & -38.09 &  	MBR178  & 4.015 & -75.60 & 289.81 & -36.27 &  	MBR222  & 3.833 & -66.36 & 280.58 & -42.19 \\
MBR135  & 3.636 & -76.23 & 291.50 & -36.95 &  	MBR179  & 2.150 & -67.60 & 292.06 & -47.86 &  	MBR223  & 3.833 & -65.13 & 279.17 & -42.84 \\
MBR136  & 3.589 & -70.68 & 286.29 & -40.74 &  	MBR180  & 2.150 & -66.37 & 291.24 & -48.96 &  	MBR224  & 3.833 & -63.90 & 277.72 & -43.48 \\
MBR137  & 3.599 & -71.91 & 287.50 & -39.91 &  	MBR181  & 2.117 & -65.14 & 290.67 & -50.15 &  	MBR225  & 4.083 & -70.67 & 284.43 & -38.75 \\
MBR138  & 3.642 & -73.15 & 288.55 & -38.97 &  	MBR182  & 2.117 & -63.91 & 289.78 & -51.25 &  	MBR226  & 4.083 & -69.44 & 283.07 & -39.38 \\
MBR139  & 3.725 & -74.38 & 289.47 & -37.91 &  	MBR183  & 2.333 & -70.06 & 292.42 & -45.22 &  	MBR227  & 4.083 & -68.21 & 281.69 & -40.00 \\
MBR140  & 3.844 & -75.61 & 290.31 & -36.77 &  	MBR184  & 2.333 & -68.83 & 291.57 & -46.30 &  	MBR228  & 4.083 & -66.98 & 280.28 & -40.60 \\
MBR141  & 3.838 & -71.30 & 285.96 & -39.40 &  	MBR185  & 2.333 & -67.59 & 290.69 & -47.37 &  	MBR229  & 4.083 & -65.75 & 278.85 & -41.18 \\
MBR142  & 3.864 & -72.53 & 287.15 & -38.57 &  	MBR186  & 2.333 & -66.36 & 289.77 & -48.44 &  	MBR230  & 4.083 & -64.52 & 277.39 & -41.75 \\
MBR143  & 3.926 & -73.76 & 288.21 & -37.63 &  	MBR187  & 2.333 & -65.13 & 288.81 & -49.50 &  	MBR231  & 4.083 & -63.29 & 275.91 & -42.30 \\
[0.5ex]
\hline
\\[-2.5ex]
\hline
\\[-1.5ex]
SMC729  & 1.392 & -68.83 & 298.61 & -48.03 &   SMC739  & 1.521 & -74.37 & 299.31 & -42.48 &   SMC816  & 1.506 & -64.52 & 296.14 & -52.09 \\
SMC730  & 1.378 & -70.06 & 299.05 & -46.83 &   SMC740  & 1.475 & -75.60 & 299.88 & -41.30 &   SMC817  & 1.481 & -63.29 & 295.91 & -53.32 \\
SMC731  & 1.370 & -71.29 & 299.42 & -45.63 &   SMC808  & 1.359 & -67.60 & 298.55 & -49.27 &   SMC824  & 1.144 & -79.91 & 301.99 & -37.19 \\
SMC732  & 1.329 & -72.52 & 299.96 & -44.44 &   SMC809  & 1.338 & -66.37 & 298.39 & -50.50 &   SMC825  & 1.144 & -78.67 & 301.85 & -38.42 \\
SMC733  & 1.276 & -73.76 & 300.52 & -43.25 &   SMC810  & 1.320 & -65.14 & 298.23 & -51.73 &   SMC826  & 1.144 & -77.44 & 301.72 & -39.64 \\
SMC734  & 1.215 & -74.99 & 301.06 & -42.06 &   SMC811  & 1.303 & -63.91 & 298.05 & -52.97 &   SMC827  & 1.560 & -79.29 & 300.47 & -37.64 \\
SMC735  & 1.143 & -76.22 & 301.58 & -40.86 &   SMC812  & 1.627 & -69.45 & 297.00 & -47.15 &   SMC828  & 1.520 & -78.06 & 300.30 & -38.86 \\
SMC736  & 1.619 & -70.67 & 297.52 & -45.97 &   SMC813  & 1.595 & -68.22 & 296.78 & -48.38 &   SMC829  & 1.487 & -76.83 & 300.13 & -40.09 \\
SMC737  & 1.594 & -71.91 & 298.12 & -44.81 &   SMC814  & 1.562 & -66.98 & 296.57 & -49.62 &           &       &        &        &        \\
SMC738  & 1.560 & -73.14 & 298.73 & -43.64 &   SMC815  & 1.532 & -65.75 & 296.36 & -50.85 &           &       &        &        &        \\
[0.5ex]
\hline
\\[-2.5ex]
\hline
\\[-1.5ex]
LMC535  & 4.506 & -73.62 & 286.47 & -35.62 &   LMC548  & 4.533 & -66.85 & 278.67 & -38.26 &   LMC619  & 4.333 & -65.13 & 277.23 & -40.05 \\
LMC536  & 4.557 & -72.38 & 284.99 & -35.96 &   LMC549  & 4.568 & -65.62 & 277.11 & -38.50 &   LMC620  & 4.250 & -64.00 & 276.15 & -40.99 \\
LMC537  & 4.602 & -71.15 & 283.49 & -36.27 &   LMC586  & 4.785 & -63.77 & 274.34 & -37.74 &   LMC621  & 4.422 & -63.77 & 275.30 & -40.03 \\
LMC538  & 4.641 & -69.92 & 281.99 & -36.56 &   LMC587  & 4.806 & -62.54 & 272.78 & -37.92 &   LMC622  & 4.456 & -62.54 & 273.69 & -40.24 \\
LMC539  & 4.677 & -68.69 & 280.47 & -36.84 &   LMC588  & 4.825 & -61.31 & 271.22 & -38.08 &   LMC623  & 4.489 & -61.31 & 272.06 & -40.41 \\
LMC540  & 4.708 & -67.46 & 278.95 & -37.09 &   LMC589  & 4.600 & -64.39 & 275.54 & -38.72 &   LMC624  & 4.433 & -79.77 & 293.13 & -32.85 \\
LMC541  & 4.736 & -66.23 & 277.42 & -37.33 &   LMC590  & 4.629 & -63.16 & 273.96 & -38.92 &   LMC625  & 4.433 & -78.54 & 291.87 & -33.49 \\
LMC542  & 4.761 & -65.00 & 275.88 & -37.55 &   LMC591  & 4.655 & -61.92 & 272.38 & -39.10 &   LMC626  & 4.433 & -77.31 & 290.59 & -34.11 \\
LMC543  & 4.287 & -73.00 & 286.35 & -36.74 &   LMC521  & 4.216 & -74.23 & 287.86 & -36.37 &   LMC627  & 4.433 & -76.08 & 289.30 & -34.72 \\
LMC544  & 4.349 & -71.77 & 284.84 & -37.09 &   LMC522  & 4.447 & -74.85 & 287.95 & -35.27 &   LMC628  & 4.733 & -79.15 & 292.02 & -32.43 \\
LMC545  & 4.403 & -70.54 & 283.31 & -37.42 &   LMC616  & 4.300 & -68.83 & 281.66 & -38.66 &   LMC629  & 4.733 & -77.92 & 290.70 & -32.97 \\
LMC546  & 4.452 & -69.31 & 281.77 & -37.72 &   LMC617  & 4.300 & -67.59 & 280.24 & -39.20 &   LMC630  & 4.733 & -76.69 & 289.37 & -33.50 \\
LMC547  & 4.495 & -68.08 & 280.23 & -38.00 &   LMC618  & 4.333 & -66.36 & 278.70 & -39.55 &   LMC631  & 4.733 & -75.46 & 288.03 & -34.01 \\
[0.5ex]
\hline
\\[-2.5ex]
\hline
\\[-1.5ex]
LMC692  & 5.187 & -51.47 & 258.51 & -36.26 &  LMC694  & 5.194 & -49.00 & 255.46 & -36.19 &  LMC702  & 5.317 & -50.85 & 257.76 & -35.04 \\
LMC693  & 5.191 & -50.23 & 256.98 & -36.23 &  LMC701  & 5.317 & -52.08 & 259.26 & -35.06 &  LMC703  & 5.317 & -49.62 & 256.26 & -35.01 \\
[-0.9ex]
\enddata
\end{deluxetable*}